\documentclass[fleqn,usenatbib]{mnras}

\usepackage{newtxtext,newtxmath}

\usepackage[T1]{fontenc}
\usepackage{ae,aecompl}

\usepackage{graphicx}	
\usepackage{amsmath}	
\usepackage{amssymb}	
\usepackage{xcolor}
\usepackage{soul}

\newcommand{\LLsum}{\sum_{-s}^{+s} \! \sum_{n=0}^{\nu \leq \nu_{\rm max}} \!}

\title[Magnetised hybrid stars]{Magnetised hybrid stars: effects of slow and rapid phase transitions at the quark-hadron interface}

\author[M. Mariani et al.]
{Mauro Mariani,$^{1,2}$\thanks{E-mail: mmariani@fcaglp.unlp.edu.ar (MM)}
Milva G. Orsaria,$^{1,2}$ 
Ignacio F. Ranea-Sandoval$^{1,2}$
\newauthor
and Germ\'an Lugones$^{3}$
\\
$^{1}$ Grupo de Gravitaci\'on, Astrof\'isica y Cosmolog\'ia, Facultad de Ciencias Astron{\'o}micas y Geof{\'i}sicas, Universidad Nacional de La Plata,\\
Paseo del Bosque S/N, 1900, Argentina\\
$^{2}$ CONICET, Godoy Cruz 2290, 1425 Buenos Aires, Argentina\\
$^{3}$ Universidade Federal do ABC, Centro de Ci\^encias Naturais e Humanas, Avenida dos Estados 5001- Bang\'u, CEP 09210-580,\\ Santo Andr\'e, SP, Brazil.}

\date{Accepted 2019 August 26. Received 2019 August 20; in original form 2019 June 12}

\pubyear{2019}

\begin{document}
\label{firstpage}
\pagerange{\pageref{firstpage}--\pageref{lastpage}}
\maketitle


\begin{abstract}
We  study the influence of strong magnetic fields in hybrid stars, composed by hadrons and a pure quark matter core, and analyse their structure and stability as well as some possible evolution channels due to the magnetic field decay. Using an ad-hoc parametrisation of the magnetic field  strength and taking into account Landau-quantization effects in matter, we calculate hybrid magnetised equations of state and some associated quantities, such as particle abundances and matter magnetisation, for different sets of parameters and different magnetic field strengths. Moreover, we compute the magnetised stable stellar configurations, the mass versus radius and the gravitational mass versus central energy density relationships, the gravitational mass versus baryon mass diagram, and the tidal deformability. Our results are in agreement with both, the $\sim 2M_\odot$ pulsars and the data obtained from GW170817. In addition, we study the stability of stellar configurations assuming that slow and rapid phase transitions occur at  the sharp hadron-quark interface. We find that, unlike in the rapid transition scenario, where $\partial M/\partial \epsilon_c < 0$ is a sufficient condition for instability, in the slow transition scenario there exists a connected extended stable branch beyond the maximum mass star, for which $\partial M/\partial \epsilon_c < 0$. Finally, analysing the gravitational mass versus baryon mass relationship, we have calculated the energy released in transitions between stable stellar configurations. We find that the inclusion of the magnetic field and the existence of new stable branches allows the possibility of new channels of transitions that fulfil the energy requirements to explain Gamma Ray Bursts.
\end{abstract}



\section{Introduction}
\label{intro}

The theoretical study of neutron stars (NS) considering the role of their intrinsic magnetic fields (MF) has become highly relevant since the discover of magnetars. Magnetars are apparently isolated NSs whose main properties can be explained by the presence of  extremely strong magnetic fields, up to $10^{15}$~G, at their surfaces \citep{Kaspi}. In general, it is believed that at least $10 \%$ of the young neutron-star population experience a magnetar phase during a short period of their evolution ($\sim 10^4$~years). During their lives they have intense activity, including persistent $X$-ray emissions, short bursts, large outbursts and giant flares which are believed to be powered by the decay of huge magnetic fields which drive changes in the crust  and the magnetosphere \citep{Duncan}.  Historically, magnetars have been associated with soft gamma repeaters (SGRs) and anomalous X-ray pulsars (AXPs), but some low magnetic field SGRs have also been found \citep{Rea1,Rea2}, as well as high surface $B$-field objects that behave as rotation powered pulsars with some occasional magnetar-like outbursts (see \cite{Archibald} and references therein).

From a model-theoretical point of view, magnetars are magnetised NSs ($M \sim 2~M_{\odot}$, $R \sim 15 \, \mathrm{km}$ \citep{Glendenning1}) with an external solid crust and an internal extremely dense core. The composition of the core is not yet fully understood since the matter density may be as large as several times the nuclear saturation density, $n_0 \approx 0.16$~fm$^{-3}$, making NSs the most dense objects in the universe. In this context, several works have analysed the role of exotic matter matter in the inner core of these objects, such as hyperons or deconfined quark matter. In any case, the equation of state (EoS) of matter should be stiff enough to allow for $2~M_{\odot}$ NSs in order to agree with the detection of the pulsars PSR J1614-2230 and PSR J0348+0432, whose masses have been very accurately determined \citep{Demorest,Antoniadis,Arzoumanian_2018}.

Moreover, the recent direct detection of gravitational waves from the NS merger GW170817 \citep{Abbott1} and its electromagnetic counterparts GRB170817A and AT2017gfo \citep{Abbott2}, imposed a new set of constrains not only on the masses but also on the radii of the coalescing objects. The upper limit on the tidal deformability of the binary system, $\tilde{\Lambda} < 800$ \citep{Abbott1}, has restricted the radius of a NS with a mass of $1.4$~$M_\odot$ to $R_{1.4} \lesssim 13.76 \, \mathrm{km}$ \citep{Raithel,Annala,Malik,Most,Fattoyev}. In summary, although the EoS of extremely dense matter is still uncertain, new multi-messenger observations provide tight constraints for current models. 

Among the possible internal compositions of compact objects, the so-called hybrid stars (HS) composed by a quark matter core and external hadronic layers have attracted much attention recently. In this work, we will analyse several properties of  hybrid stars with strong magnetic fields at their interiors.  In this context, it is necessary to model the boundary between the hadron and the quark phases. However, the very nature of such interface is still uncertain. It has been speculated that hybrid stars may contain a mixed hadron-quark phase in their interiors (see \cite{Orsaria} and references therein). In such a phase it is assumed that the electric charge is zero globally but not locally, and therefore charged hadron and quark matter may share a common lepton background, leading to a quark-hadron mixture extending over a wide density region of the star \citep{Glendenning2}. The mixed phase entails a smooth variation of the energy density, leading in turn to a continuous density profile along the star. On the other hand, the quark matter core could be separated from the external hadronic layers of the star by a sharp interface with a density jump across which thermal, mechanical and chemical equilibrium is maintained. Whether the quark-hadron interface is actually a sharp discontinuity or a wide mixed region depends crucially on the amount of electrostatic and surface energy needed for the formation of diverse geometric structures of one phase embedded in the other all along the mixed phase \citep{Voskresensky,Endo,Yasutake}. If the energy cost of Coulomb and surface effects exceeds the gain in bulk energy, the scenario involving a sharp interface turns out to be favourable. However,  there are still many points to be elucidated before arriving to a conclusive description of quark-hadron coexistence. In particular, model calculations of the surface tension span a wide range of values (see \cite{Lugones3,Lugones2} and references therein), with results both above and below the critical value that favours a sharp interface. In this work, we assume that the interface between hadrons and quarks is a sharp discontinuity.

In order to understand the structure of magnetic fields in the interior of compact stars it would be necessary to solve the general relativistic magneto-hydrodynamic equations to determine the MF direction and strength at any point of the star. However, finding magnetised equilibrium NSs models in General Relativity is a very complex theoretical problem, which has not been fully solved yet. One of the difficulties is the non-linear nature of Einstein's equations for the metric. For realistic MF configurations including poloidal and toroidal components and if rotation is included,  many metric terms must be retained and a large set of coupled elliptic partial differential equations has to be solved by means of numerical methods (see e.g. \cite{Pili} and references therein). Another important aspect is related to the way the MF evolves and dissipates in the internal layers of the NS; not only in the crust but also in the core, which has an uncertain composition. Since the ionic lattice of the crust is static, the MF evolves exclusively through Ohmic decay and Hall drift. However, the MF evolution of the core is much more complex and uncertain because of the possible presence of several particle species, the effect of weak interactions ($\beta$-equilibrium) and particle diffusion, the possible existence of superfluidity/superconductivity, etc. (see a thorough list of references in \cite{Gusakov}). Additionally, it is still under debate to what extent magneto-hydrodynamic equilibrium in NSs are arbitrary for given EOSs, with or without axisymmetry \citep{Glampedakis,Gusakov}. 

In this work, we would like to avoid the complexities associated with the multidimensional nature of the MF geometry and focus on another relevant aspect of strong magnetic fields: its influence on the hadron-quark equations of state and its role on the dynamic stability of hybrid stars assuming that phase conversions at the quark-hadron interface are either slow or rapid. To this end, we will assume, as a working hypothesis, that the structure of a strongly magnetised NS can be reasonably described by means of the standard spherically symmetric Tolman-Oppenheimer-Volkoff (TOV) stellar structure equations. This hypothesis is based on the following qualitative  assumptions. Firstly, during the hot proto-NS phase, differential rotation would create strong toroidal MFs inside the NS \citep{Bonano,Naso,Frieben}. As a result, realistic models of magnetised NSs require the simultaneous presence of both poloidal and toroidal MF components \citep{Ciolfi}. Secondly, it is known that purely toroidal MFs make the NS prolate, while purely poloidal MFs tend to make it oblate. If both the toroidal and poloidal components are of the same order, we may expect that oblateness and prolateness cancel out approximately, leading to stars close to the spherical symmetry. Thirdly, the inclusion of the MF in the EoS is usually done by assuming that $B$ is \textit{locally} uniform, which leads to different values of the pressure parallel and transverse to the \textit{local} direction of $B$ ($P_{\perp}$ and  $P_{\parallel}$). Since on macroscopic scales the MF is supposed to be  a geometrically complex combination of poloidal and toroidal fields, with the MF direction changing disorderly from one point to the other,  we will average $P_{\perp}$ and  $P_{\parallel}$ in order to obtain a \textit{locally}  effective isotropic pressure (see \cite{Bednarek,Flores} for a similar procedure). Finally, we will assume that the MF strength is a monotonically decreasing function of the distance to the NS centre. Such behaviour will be incorporated through an ad-hoc monotonic function connecting the MF strength with the baryon chemical potential. This is clearly a strong assumption about the MF behaviour inside NSs, but it has been widely used in the literature in order to assess the consequences that could be expected if ultrahigh MFs were present at magnetar cores \citep{Bandyopadhyay,Mao,Rabhi,Dexheimer}.

On the other hand, we will analyse the dynamic stability of hybrid star configurations using the formalism of small radial perturbations \citep{Chandrasekhar}. Such formalism has been adapted by \cite*{Pereira1} to take into account the role of  phase conversions in the vicinity of a sharp hadron-quark interface. Hadron-quark phase transition may give rise to density fluctuations at the interface that may have certain influence on the dynamical evolution of the system. For example, local fluctuations give rise to microscopic quark clusters formation involving a large free energy barrier between the two-phases structure, so that the process can be very slow, as in the nucleation process. Alternatively, the phase transition could be analysed through a spinodal decomposition if the formation of a sudden two-phases structure occurs by a diffusion process in which the fluctuations are long range, the system becomes thermodynamically unstable and so the transition process occurs rapidly (see \cite{malfatti2019} and references therein). One of the main results found by \cite*{Pereira1}  is that  the usual static stability condition  $\partial M / \partial \epsilon_c \ge 0$, where $\epsilon_c$ is the central density of a star whose total mass is $M$, always remains true if phase conversions are rapid but breaks down in general if they are slow. As a consequence, an additional branch of stable HS configurations is possible in the case of slow phase conversions. In this work, we will analyse in detail the structure and stability of  low-B HSs and magnetars under rapid and slow phase conversions. Besides, we will analyse the gravitational mass versus baryon mass diagram to estimate the energy released when a magnetar evolves to become a low-B object, based on the ideas developed in \cite{Bombaci1} and already presented in \cite{Mariani}.

In order to automate the calculations and integrate the whole process, we use the multi-language NeStOR code introduced in \cite{Mariani}. Originally developed to construct EoS at finite temperature and calculate the structure of proto-NS, we adapted the code to run the magnetised EoS codes and implemented a procedure to detect the change of stability of radial oscillation modes.

The paper is organised as follows. In Section~\ref{eos}, we describe the adopted ansatz for the magnetic field strength and give the thermodynamic expressions for the hadronic and quark phases of the EoS. We also present the conditions for the construction of a hybrid star with a sharp transition and discuss our approach to the anisotropy of the MF. In Section~\ref{structure}, we present the stellar structure equations and the formalism employed to analyse the dynamic stability of stellar configurations when slow and rapid phase conversions are present at the quark-hadron interface. We also show our results for magnetised hybrid stars, considering the observational constrains. A summary of the work, discussion about the astrophysical implications of our results and conclusions are provided in Section~\ref{conclus}. In Appendix \ref{appe}, some issues related to the electromagnetic units are discussed and clarified.


\section{Hybrid magnetised EoS}
\label{eos}

\subsection{Ansatz for the MF strength}  
\label{mageos}

In order to include the effect of the variation of the MF strength in a magnetised NS from its centre to its surface, we consider a MF in the $z$-direction depending on the chemical potential, $\mu_b$, in the following form:
\begin{equation} \label{param}
	B(\mu_b) = B_{\text{min}}+B_{\text{max}} \left[ 1-\mathrm{e}^{\beta\frac{(\mu_b-m_n)^\alpha}{m_n}} \right] \,
\end{equation}
where $\alpha=2.5$ and $\beta = -4.08 \times 10^{-4}$ are fixed parameters and $m_n$ is the nucleon mass \citep{Dexheimer}. 

The functional form and the values of the parameters were chosen to reproduce the values of a MF parametrised in terms of the density from \cite{Dexheimer, Rabhi, Mao}. Notice that this parametrisation is independent of the EoS and generates no discontinuity at the quark-hadron interface of the hybrid star. As it will be seen in following sections, one of the consequences of this ansatz is that only stars with very high central density reach values close to $B_{\text{max}}$ for its central MF, as it has been suggested in \cite{Dexheimer}. In this work, we consider two representative cases for this parametrisation: the \mbox{low-$B$ HS} case, with $B_{\text{min}}=1.0 \times 10^{13}$~G and $B_{\text{max}}=1.0\times10^{15}$~G, and the magnetar case, with $B_{\text{min}}=1.0\times10^{15}$~G and $B_{\text{max}}=3.0\times 10^{18}$~G.

It should be noted that we use the natural units system, in which the value of fundamental constants, $c=\hbar=1$. More details about the electromagnetic and MF units are given in Appendix \ref{appe}.

\subsection{Landau levels}
\label{sec:landau}

In the presence of a MF, the  motion of electrically charged particles  is quantized into Landau levels in the direction perpendicular to the local direction of $B$. We assume that the MF points in the $z$-direction, locally, and particles have electric charge,  $q$, and spin, $s$. As a result of Landau quantization, a quantum number $\nu$ arises, which depends on the angular quantum number, $n$, and the spin projection of the particle, $\sigma = -s,...,+s$, \citep{Landau}
\begin{equation} 
\label{landaunu}
\nu = n + \frac{1}{2} - \frac{g}{2} \text{sgn}(q) \, \sigma \, ,
\end{equation}
where $\text{sgn}(x)$ is the sign function. The $g$-factor is $g=2$ for spin $1/2$ particles and $g=2/3$ for spin $3/2$ particles  \citep{Belinfante,Torres}. The transverse momentum of a particle is given by 
\begin{equation}
 	k_\perp^2 = 2 \nu |q| B \, ,
\end{equation}
and the energy spectrum by
\begin{equation}
	E = \sqrt{k_z^2 + \bar{m}^2(\nu)} \, ,
\end{equation}
where $\bar{m}^2(\nu) = m^2 + 2 \nu |q| B$. As a consequence, the $z$-component of the Fermi momentum reads,
\begin{equation}
	k_{F,z}(\nu) = \sqrt{\mu^2 - 2 \nu |q| B(\mu) - m^2} \, .
\end{equation}
In order to keep $k_{F,z}$ real valued, we must impose $\nu \le\nu_{\rm max}$, with
\begin{equation}
	\nu_{\rm max} = \left\lfloor \frac{\mu^2 - m^2}{2|q|B} \right\rfloor ,
\label{numax} 
\end{equation}
where $\lfloor x \rfloor$ is the largest integer less than or equal to $x$. 

To implement the effect of Landau quantization in the thermodynamic quantities,  the following replacement must be done in all thermodynamic integrals:
\begin{equation}
\frac{1}{(2 \pi)^{3}} \int \cdots d^{3} k \longrightarrow \frac{\left|q B\right|}{2 \pi^{2}} 
\sum_{-s}^{+s} \sum_{n=0}^{\nu \leq v_{\max }}
\int_{0}^{k_{F,z}(\nu)} \cdots d k_{z} . 
\end{equation}

\subsection{Average of the pressure anisotropy}
\label{sec:average}

The energy-momentum tensor of matter in the presence of a MF can be split in two components,
\begin{equation}
T_{\mu\nu} = T_{\mu\nu}^{\text{matter}} + T_{\mu\nu}^{\text{MF}} \, ,
\end{equation}
where the first term represents the contribution of matter and the second term is the purely electromagnetic part.

As it has been mentioned before, from a microscopic point of view the MF can be seen as being locally uniform and pointing in the $z$-direction. As a consequence, an asymmetry emerges between the pressure $P_\parallel$ along the local direction of $B$, and the pressure $P_\perp$ in the transverse direction. Thus, $T^{\text{matter}}_{\mu \nu}$ can be written as 
\begin{equation}
 	T^{\text{matter}}_{\mu \nu} = \text{diag}(\epsilon, P_\perp, P_\perp,  P_\parallel) \, ,
\end{equation}
where $\epsilon$ is the energy density and $P_\perp$, $P_\parallel$ are given by
\begin{align}
P_\parallel &= - \Omega \, , \nonumber \\
P_\perp &= - \Omega - \mathcal{M}B \, ,
\label{pppper}
\end{align}
which are related via the total matter magnetisation \citep{Gonzalez, Blanford},
\begin{equation}
    \mathcal{M}= - \partial \Omega / \partial B \rvert_{\mu_B}\, .
\label{magmatter}
\end{equation}
In equations \eqref{pppper} and \eqref{magmatter}, $\Omega$ represents the grand canonical potential of the system. In addition, there is another contribution to the anisotropy of the system coming from the pure magnetic field terms, 
\begin{equation}
 	 T^{\text{MF}}_{\mu \nu} = \text{diag}(B^2/2,  B^2/2, B^2/2,  -B^2/2) \, .
\end{equation}
Thus, the total energy-momentum tensor reads,
\begin{align}
T_{\mu \nu} &=T^{\text{matter}}_{\mu \nu}+T^{\text{MF}}_{\mu \nu} \nonumber \\ 
&=\text{diag}(\epsilon+B^2/2, P_\perp+B^2/2, P_\perp+B^2/2,  P_\parallel-B^2/2) \, .
\end{align}

It will be shown in Section~\ref{HEoS_phase}, that the anisotropy coming from the magnetisation in  Eq.~\eqref{magmatter} does not contribute significantly to the EoS and can be neglected since it does not affect the global properties of the star. However, we will see that the anisotropies of $T_{\mu \nu}$  coming from the pure magnetic field terms may be non-negligible in some high-density layers of  the NSs permeated by extremely large MFs (see e.g. Fig.~\ref{fig:pppt1}). To ensure the results obtained from the TOV equations are still valid when such anisotropies are present, we assume that in the NS core the MF is a complex combination of poloidal and toroidal fields, with the MF direction changing disorderly from one point to the other. In such circumstances, we will average the spatial components of $T_{\mu \nu}$ in order to obtain a \textit{locally}  effective isotropic pressure \citep{Bednarek,Flores}:
\begin{equation}
	P=\frac{T_{1 1}+T_{2 2}+T_{3 3}}{3}=\frac{2 P_\perp+P_\parallel}{3} + \frac{B^2}{6}=P_{\text{matter}}+P_{\text{MF}} \, .
\label{pressure_prescription}	
\end{equation}
With this prescription, we are able to construct the EoS of the system, $P=P(\epsilon)$, which will allow us to obtain families of magnetised hybrid stars using the TOV equations.

The specific expressions for the  thermodynamic quantities of matter depend on the particular model of each phase, which will be described in the next subsection.

\subsection{Magnetised EoS phases}

\subsubsection{Sub-nuclear matter}

To model the magnetised matter at sub-nuclear densities in the crust of the HS, we use an EoS based on the BPS \citep{Baym1} model with a MF of $10^{15}$~G \citep{Lai}. In such work, the authors conclude that, for densities typically found in the outer core of NSs, the results for magnetised crust EoS differs from the non-magnetic crust EoS only at densities below  $\epsilon = 0.2124~\mathrm{MeV/fm}^3 \approx 4 \times 10^{12}~\mathrm{g/cm}^3$. Thus, we consider this magnetised crust EoS up to $\sim 4 \times 10^{12}~\mathrm{g/cm}^3$ and the non-magnetic BPS-BBP EoS for higher density values of the crust up to $\sim 10^{14}~\mathrm{g/cm}^3$ \citep{Baym1,Baym2}.

\subsubsection{Hadron matter}

To model magnetised hadron matter in the outer core of the HS, we use the GM1L density-dependent relativistic mean-field (RMF) parametrisation in which the interactions between baryons are described by the exchange of scalar ($\sigma$), vector ($\omega$), and isovector ($\rho$) mesons. We consider the presence of the baryon octet (neutron, proton, $\Sigma^-$, $\Sigma^0$, $\Sigma^+$, $\Lambda$, $\Xi^-$ and $\Xi^0$) and the four $\Delta$-resonances ($\Delta^-$, $\Delta^0$, $\Delta^+$, $\Delta^{++}$). Although this model has been used before to describe neutron star matter \citep{Spinella,Ranea:2019eoh,malfatti2019}, we extend it including MF effects.

The functional form of the density dependent coupling constants related with the $\rho$ meson are given by \cite{Typel2},
\begin{equation}
g_{\rho b}(n_B) = g_{\rho b}(n_0) \, \mathrm{exp} \left[ -a_{\rho} \left( \frac{n_B}{n_0} - 1 \right) \right] \, ,
\end{equation}
where $g_{\rho b}(n_B)$ is the density dependent meson-baryon coupling constant and $n_B = \sum_b n_b$ is the total baryon number density.  The parameter $a_{\rho}$ is fixed by the binding energies, charge and diffraction radii, spin-orbit splittings, and the neutron skin thickness of finite nuclei \citep{Typel1}. 

\begin{table}
\begin{center}
\begin{tabular}{|c|c|}
\hline 
$~~$Saturation Property$~~$ & $~~$GM1L$~~$\\ \hline
$n_0$  (fm$^{-3}$)    & 0.153      \\
$E_0$  (MeV)          & $-16.30$    \\
$K_0$  (MeV)          & 300.0       \\
$m^*/m_N$             & 0.70        \\
$J$    (MeV)          & 32.5          \\
$L_0$  (MeV)          & 55.0         \\
$-U_N$ (MeV)        &65.5          \\ \hline
\end{tabular}
  \caption{Properties of nuclear matter at saturation density computed for the GM1L parametrisation \protect \citep{Spinella}.}
\label{table:properties}
\end{center}
\end{table}

The meson-hyperon coupling constants have been determined following the Nijmegen extended soft core (ESC08) model \citep{Rijken}. The
relative isovector meson-hyperon coupling constants were scaled with the hyperon isospin and, for $\Delta-$resonances, $x_{{\sigma}{\Delta}} = x_{{\omega} {\Delta}} = 1.1$ and $x_{{\rho}{\Delta}} = 1.0$, where $x_{ i H} = g_{i H}/g_{i N}$ was used (see \cite{Spinella} for details).

\begin{table}
\begin{center}
\begin{tabular}{|c|c|}
\hline 
$~~$Parameter$~~$ & $~~$GM1L$~~$\\ \hline
$m_{\sigma}$  (GeV)    & 0.5500         \\
$m_{\omega}$  (GeV)          &0.7830   \\
$m_{\rho}$  (GeV)          & 0.7700          \\
$g_{\sigma N}$             & 9.5722       \\
$g_{\omega N}$            & 10.6180          \\
$g_{\rho N}$            & 8.9830         \\
$b_{\sigma}$         &0.0029                     \\
$c_{\sigma}$         &$- 0.0011$              \\
$a_{\rho}$         &0.3898        \\ \hline
\end{tabular}
  \caption{Parameters of the GM1L parametrisation that lead to the properties of symmetric nuclear matter at saturation density given
    in Table \ref{table:properties}.}
\label{table:parametrisations}
\end{center}
\end{table}

The saturation properties including the nuclear saturation density, $n_0$, energy per nucleon, $E_0$, nuclear incompressibility, $K_0$, effective nucleon mass, $m^*/m_N$, asymmetry energy, $J$, asymmetry energy slope, $L_0$, and nucleon potential, $U_N$, and the parameters of the GM1L model used here are listed in Tables \ref{table:properties} and \ref{table:parametrisations}. The values for baryon masses and $\Delta-$resonances are taken from  \cite{Spinella}.

The meson mean-field equations are given by
\begin{align}
m_{\sigma}^2 \bar{\sigma} &= \sum_{b} g_{\sigma b} n_b^s - b_{\sigma} m_N g_{\sigma N} (g_{\sigma N}\bar{\sigma})^2 \nonumber \\
& - c_{\sigma} g_{\sigma N} (g_{\sigma N} \bar{\sigma})^3 \, , \nonumber \\
m_{\omega}^2 \bar{\omega} &= \sum_{b} g_{\omega b} n_{b}^v \, , \\
m_{\rho}^2\bar{\rho} &= \sum_{b}g_{\rho b}(n_B)I_{3b} n_{b}^v \, , \nonumber
\end{align}
where $I_{3b}$ is the 3-component of isospin,  $n_{b}^s$ and $n_{b}$ are the scalar and particle number densities for each baryon $b$,  $\bar{\sigma}$, $\bar{\omega}$ and $\bar{\rho}$ are the meson mean fields in the RMF approximation, and $m_N$ is the neutron mass. 

For charged baryons in an external magnetic field, $B$, the scalar and vector densities are given by
\begin{equation}
n_{c,b}^s = \frac{|q_b| B \, m_b^*}{2 \pi^2} \LLsum {\rm{ln}} \bigg\lvert \frac{k_{z, \rm{F}}^{b} + E_{\rm{F}}^b}{\bar{m}_{b}} \bigg\rvert,
\end{equation}
\begin{equation}
n_{c,b}^v = \frac{|q_b| B}{2 \pi^2} \LLsum k_{z,\rm{F}}^{b},
\end{equation}
where the double sum goes over the discrete angular momentum quantum number, $n$, and over the spin projection, $\sigma = -s,...,+s$, and $q_b$ is the electric charge of the baryon $b$. The effective masses, the $z$-component of the Fermi momentum, and the Fermi energy are respectively given by
\begin{align}
m_b^*&= m_b - g_{\sigma b} \bar{\sigma} \, ,\nonumber \\ 
\bar{m}_{b} &= \sqrt{m_b^{*2} + 2 \nu | q_b | B} \, ,\nonumber \\ 
k_{z,\rm{F}}^{b} &= \sqrt{E_{\rm{F}}^b - \bar{m}_b^2} \, ,\nonumber \\ 
E_{\rm{F}}^b &= \mu_b- g_{\omega b}(n) \bar{\omega} - g_{\rho b}(n)\bar{\rho} I_{3b} - \widetilde{R} \, ,
\label{eqshad}
\end{align}
where $\widetilde{R}$ is the rearrangement term given by
\begin{equation}
\widetilde{R} = \sum_b \frac{\partial g_{\rho b}(n)}{\partial n}
I_{3b} n_b^v \bar{\rho} \, ,
\label{rear}
\end{equation}
which is important for achieving thermodynamic consistency \citep{Hofmann}. The energy density for charged baryons can be expressed by
\begin{equation}
\epsilon_{c,b} = \frac{|q_b| B}{4 \pi^2} \LLsum \left( k_{z,\rm{F}}^{b} E_{\rm{F}}^b + \bar{m}_{b}^2 {\rm{ln}} \bigg\lvert \frac{k_{z,\rm{F}}^{b} + E_{\rm{F}}^b}{\bar{m}_{b}} \bigg\rvert \right) \, .
\end{equation}

For neutral baryons, the scalar and vector densities are given by
\begin{equation}
n_{n,b}^s = \frac{\gamma_s \, m_b^*}{4 \pi^2} \left( E_{\rm{F}}^b k_{\rm{F}}^b - m_b^{*2} {\rm{ln}} \bigg\lvert \frac{k_{\rm{F}}^{b} + E_{\rm{F}}^b}{{m}_{b}^*} \bigg\rvert \right) \, ,
\end{equation}
\begin{equation}
n_{n,b}^v = \frac{\gamma_s}{2 \pi^2} \frac{(k_{\rm{F}}^{b})^3}{3} \, ,
\end{equation}
where the spin degeneracy factor, $\gamma_s=2$ for spin $1/2$ particles and $\gamma_s=4$ for spin $3/2$ particles. Eqs.~(\ref{eqshad}) are also valid for neutral baryons, but with  $\bar{m_b} = m_b^*$ and using the total Fermi momentum, $k_{\rm{F}}^b = \sqrt{E_{\rm{F}}^b - m_b^{*2}}$, instead of its $z$-component. The energy density for neutral baryons is given by
\begin{equation}
\epsilon_{n,b} = \frac{\gamma_s}{4 \pi^2} \left[\frac{1}{2} (E_{\rm{F}}^b)^3 k_{\rm{F}}^b - \frac{1}{4} m_b^{*2} \left( E_{\rm{F}}^b k_{\rm{F}}^b + m_b^{*2} {\rm{ln}} \bigg\lvert \frac{k_{\rm{F}}^{b} + E_{\rm{F}}^b}{{m}_{b}^*} \bigg\rvert \right) \right] \, .
\end{equation}
The total baryon energy density is given by
\begin{align}
\epsilon_B =& \sum_b \epsilon_{b} + \frac{1}{2} \left( m_{\sigma}^2\bar{\sigma}^2 + m_{\omega}^2 \bar{\omega}^2 + m_{\rho}^2 \bar{\rho}^2 \right) + \nonumber \\
 &+ \frac{1}{3} b_{\sigma} m_N g_{\sigma N} \bar{\sigma}^3 + \frac{1}{4} c_{\sigma} m_N g_{\sigma N} \bar{\sigma}^4 \, .
\label{enden}
\end{align}

The expression for the anisotropic pressures are determined by the density energy given in Eq.~\eqref{enden}, being
\begin{equation}
P_{\parallel} = \sum_b \mu_b n_b^v - \epsilon \, ,
\label{ppareq}
\end{equation}
\begin{equation}
P_{\perp} = \sum_b \mu_b n_b^v - \epsilon - \mathcal{M} B \, ,
\label{ppereq}
\end{equation}
where the sum goes over both, neutral and charged baryons. Taking into account Eqs.~(\ref{ppareq}) and (\ref{ppereq}) and following the prescription given in Eq.~\eqref{pressure_prescription}, we obtain the EoS for magnetised hadronic matter, $P_{\text{hadron}}=P_{\text{hadron}}(\epsilon)$.

As we have already established, besides the baryon octet, we include the four $\Delta$-resonances in the hadronic phase. Unlike all the other particles we consider in this work, $\Delta$ particles have spin $s=3/2$. This property modifies the values taken by the quantum number $\nu$ and, consequently, the way to sum over the Landau levels, which has been considered in Eq.~(\ref{landaunu}).

\subsubsection{Quark matter}
\label{sec:fcm}

For the inner core of the HS, we consider the presence of magnetised deconfined $u$, $d$ and $s$ quark matter at zero temperature described through the Field Correlator Method (FCM). The FCM considers that single quark and gluon lines interact in non-perturbative way, known as single line approximation, using strength field correlators, with vacuum fields (gluon and quark condensates) \citep{Simonov2}. In this approximation, the partition function factorises into a product of one-gluon and one-quark (antiquark) contributions, and then the corresponding thermodynamic potential can be calculated at finite temperature \citep{Simonov1}. 

The FCM quark model is characterised by two quantities: the gluon condensate, $G_2$, and the large distance static $\bar{q}q$ potential, $V_1$. As we discuss in a previous work, the values of $V_1$ and $G_2$ could depend $a$ $priori$ on the temperature and baryon density (see \cite{Mariani} and references therein). At zero baryon density and finite temperature, the phenomenological values of $V_1$ and $G_2$ are constrained by lattice calculations, but these values could be very different at low temperatures and large baryon densities, as occurs in the core of magnetised HSs \citep{Burgio}. In our case, as we work under the zero temperature-high densities regime, we consider $V_1$ and $G_2$ as free constant parameters. Thus, the effects of such parameters in our model will lead to a shifted chemical potential, $\tilde{\mu} = \mu - V_1/2$, and an extra vacuum pressure term, proportional to $G_2$.

Following \cite{Strickland}, where the magnetised EoS of a Fermi gas subject to a constant magnetic field is calculated, we include the Landau levels into the FCM, obtaining the number density,
\begin{equation}
n_q = \frac{\gamma_c |q_q| B}{2\pi^2} \LLsum \, k_{z,F}^q \, ,
\label{ndense}
\end{equation}
the energy density,
\begin{equation}
\epsilon_q = \frac{\gamma_c |q_q| B}{4\pi^2} \LLsum \Biggl[ E_{\rm{F}}^q \, k_{z,F}^q + \bar{m}^2 \ln\left(\frac{ E_{\rm{F}}^q + k_{z,F}^q}{\bar{m}}\right) \Biggr] \, ,
\label{edense}
\end{equation}
and the parallel and transverse pressure components,
\begin{align}
& P_{\parallel,q} = \frac{\gamma_c |q_q| B}{4\pi^2} \LLsum \Biggl[ E_{\rm{F}}^q \, k_{z,F}^q - \bar{m}^2 \ln\left(\frac{ E_{\rm{F}}^q + k_{z,F}^q}{\bar{m}}\right) \Biggr] \, , \nonumber\\
& P_{\perp,q} = \frac{\gamma_c |q_q|^2 B^2}{2 \pi^2} \LLsum \, \nu \, \ln\left(\frac{ E_{\rm{F}}^q + k_{z,F}^q}{\bar{m}}\right) \,\label{p1} .
\end{align}
The double sum goes over the discrete angular momentum quantum number, $n$, and over the spin projection, $\sigma = -s,...,+s$, and $q_q$ is the electric charge of the considered quark and $\gamma_c$ is the degeneracy colour factor, $\gamma_c=3$. At zero temperature the gluons do not contribute to the  pressure so we do not consider them. The effective mass, the $z$-component of the Fermi momentum and the Fermi energy are respectively given by
\begin{align}
\bar{m}^2 &= m^2 + 2 \nu |q| B \, , \\
k_{z,F}^q &=  \sqrt{\tilde{\mu}_q^2 - 2 \nu |q| B - m^2} \, , \\
E_{\rm{F}}^q &=\tilde{\mu}_q \, .
\end{align}
As specified in Subsection~\ref{sec:landau}, the restriction $\nu \le \nu_{\rm max}$ comes from the fact that $k_{z,F}(\nu)$ has to remain real.

Eqs.~(\ref{ndense}), (\ref{edense}) and (\ref{p1}) satisfy the thermodynamic relationship for the grand canonical potential, $\Omega = \epsilon - \mu n = - P_\parallel$, and the canonical relationship between the pressure components, $P_\perp = P_\parallel - \mathcal{M} B$, being $\mathcal{M}$ the matter magnetisation. Hence, following the prescription given in Eq.~\eqref{pressure_prescription},
the total quark matter pressure $P_{\rm{matter}}$ reads
\begin{equation}
	P_{\rm{matter}} = \sum_{q=u,d,s} (2 P_{\perp,q}+P_{\parallel,q})/3 \, + P_{vac} \, \label{pmat} ,
\end{equation}
where, $P_{\rm{vac}}$ is the vacuum pressure given by \citep{Simonov1,Simonov2,Nefediev}
\begin{equation}
	P_{\rm{vac}} = - \frac{9}{64} G_2 \, .
\end{equation}

Therefore, using Eq.~(\ref{pmat}) and following the prescription given in the previous subsection, we obtain the quark magnetised EoS, $P_{\rm{quark}}=P_{\rm{quark}}(\epsilon)$.

\subsubsection{Leptons}

Since we work at zero temperature, we only consider electrons and muons for the leptonic contribution present in both, the hadron and quark phases. Leptons are treated as free Fermi gases with the magnetic field distribution given by Eq.~(\ref{param}) and the corresponding quantized Landau levels. The expressions for the thermodynamic quantities are the same as in the case of quark matter, taking into account the corresponding values for the lepton masses and lepton electric charges, setting the degeneracy factor $\gamma_c=1$ instead of $\gamma_c=3$ and the parameters $V_1 = G_2 = 0$.

\subsection{Hybrid EoS and phase transition}
\label{HEoS_phase}

Once the EoS for each phase is computed, it is possible to construct the hybrid magnetised EoS. We work under the formalism known as {\it Maxwell construction} which considers a sharp transition without a mixed phase region between hadron and quark matter. As it has been mentioned before, this would be the case in NSs if surface and Coulomb effects in the quark-hadron interface are large enough. In this situation, the crossing of the hadron and quark EoS in the $\mu-P$ plane defines the phase transition.

\begin{figure}
\begin{center}
	\includegraphics[width=0.90\columnwidth]{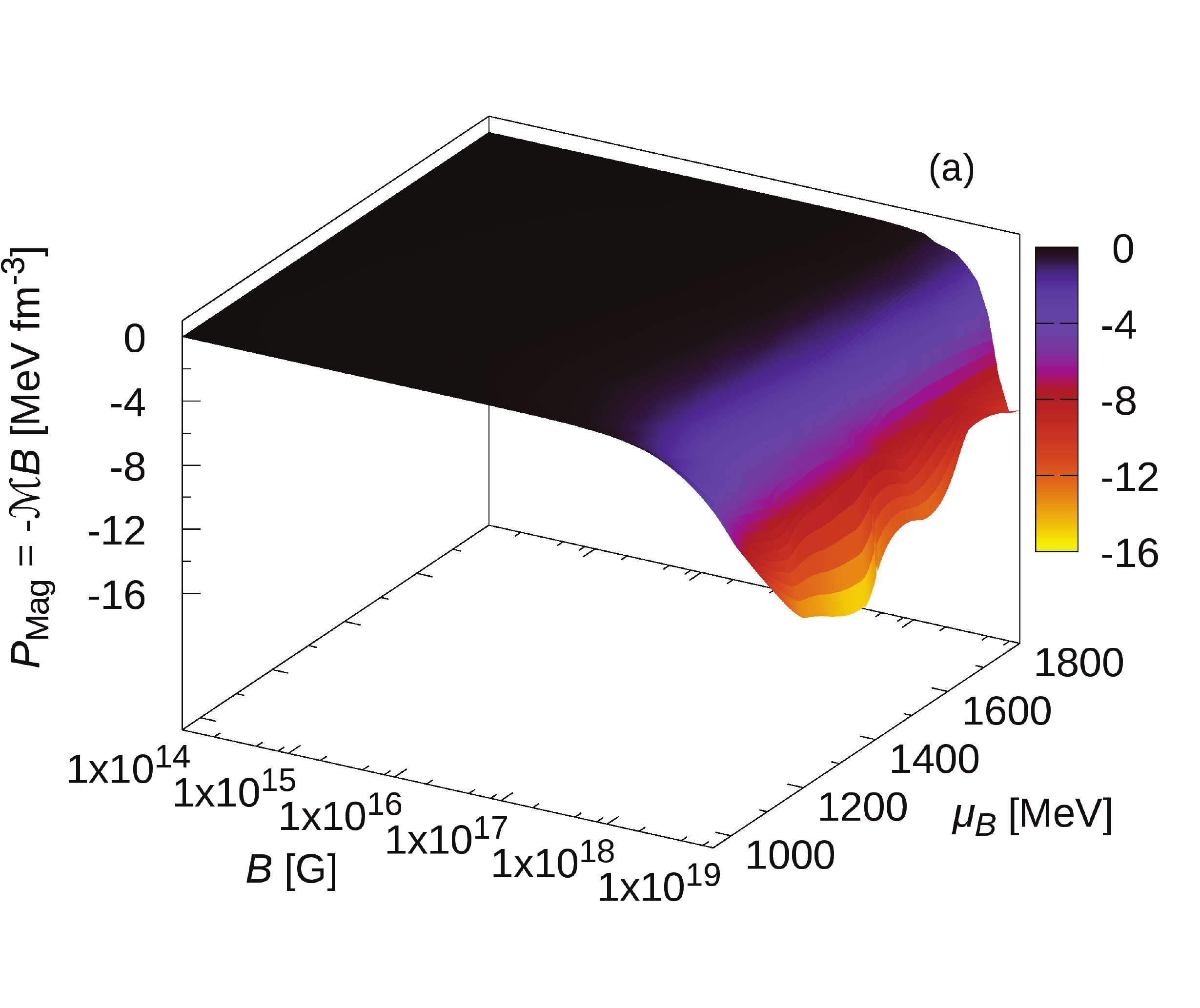}\\
	\includegraphics[width=0.90\columnwidth]{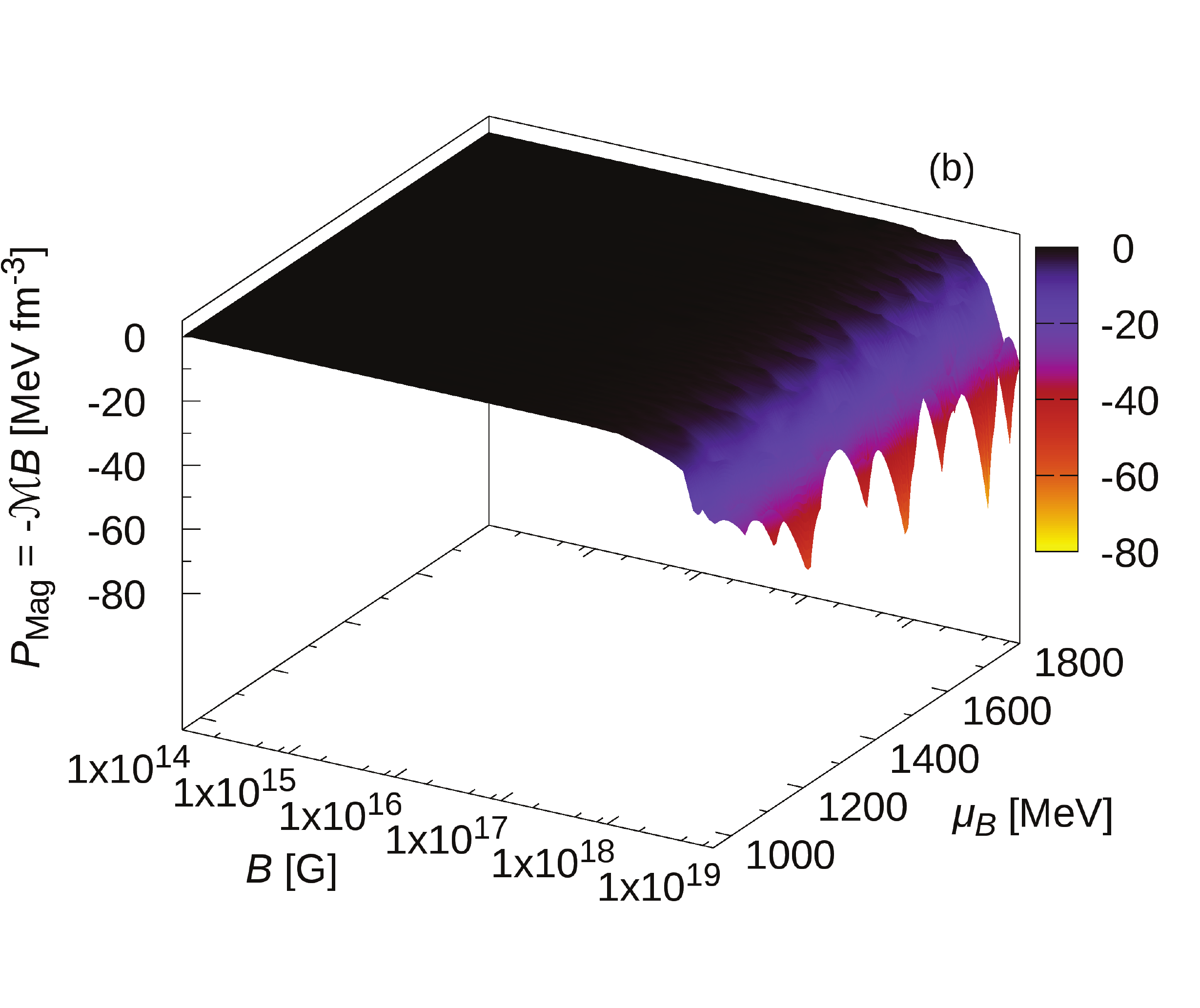}
    \caption{(Colour online) Magnetisation pressure, $P_{\rm Mag}= - \mathcal{M}B$, as a function of the MF and the baryon chemical potential. The colour bars also indicate the value of $P_{\rm Mag}$. The top (bottom) panel shows the hadronic (quark) matter case. For the hadron phase, it can be seen that only for the highest MF values the magnetisation pressure is non-negligible. At this MF order values, the phase transition has already occur so we do not consider the magnetic pressure in the hadron EoS. For the quark phase, it can be noticed the oscillatory effects of the Landau levels and that, only when the MF is extremely high, the magnetisation takes considerable values.}
    \label{fig:premag}
\end{center}    
\end{figure}

\begin{figure}
\begin{center}
	\includegraphics[width=1.0\columnwidth]{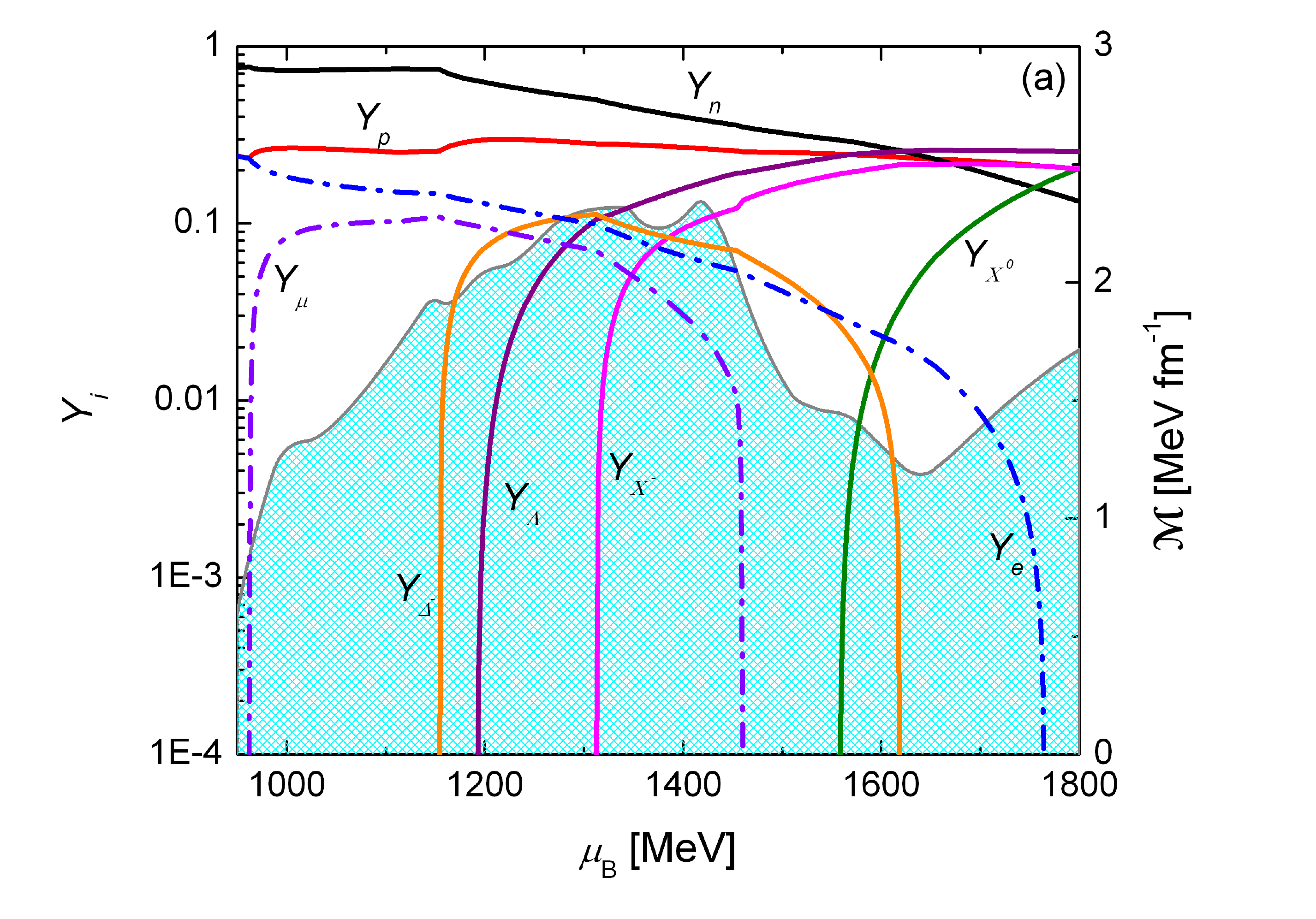}\\
	\includegraphics[width=1.0\columnwidth]{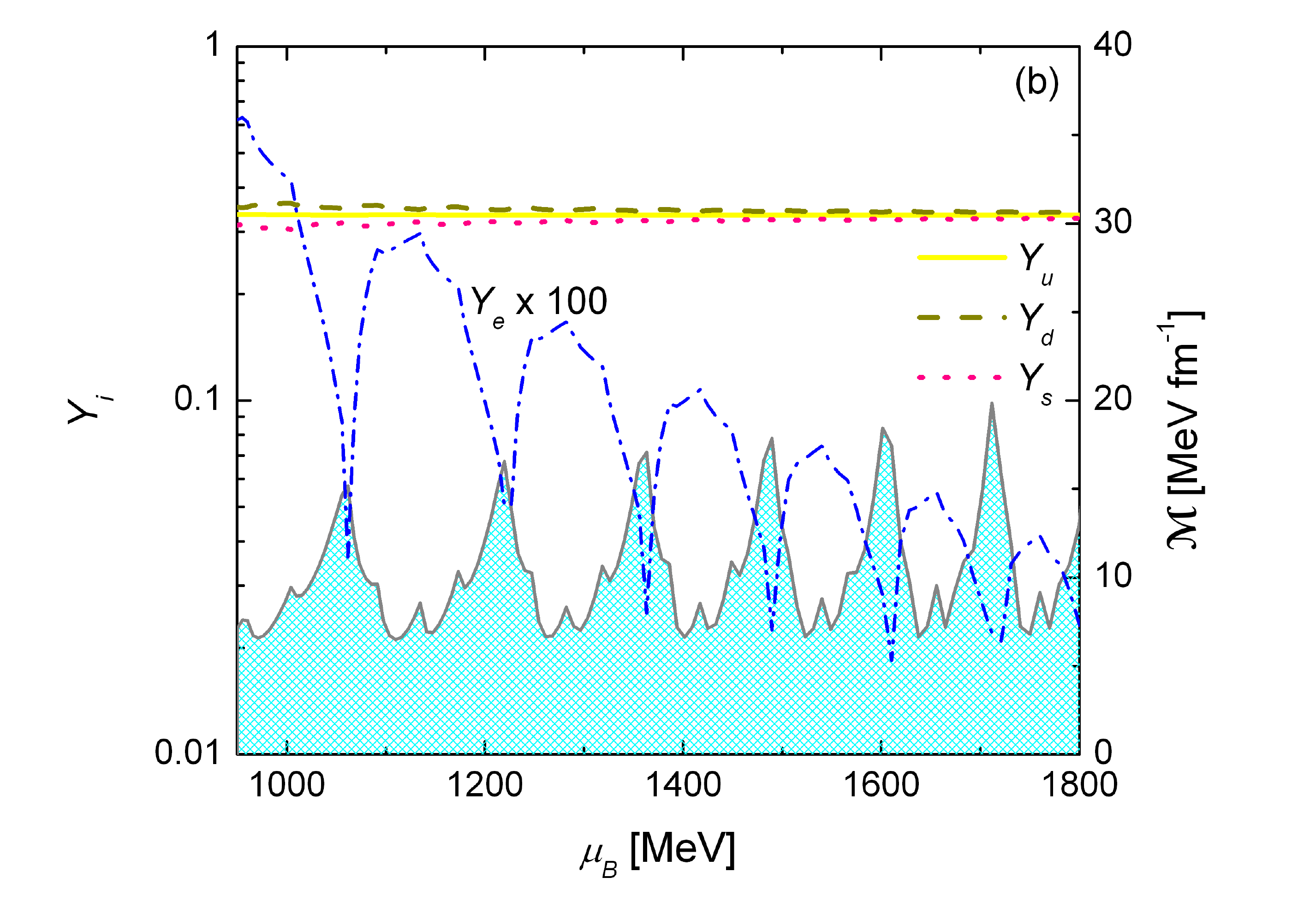}
    \caption{(Colour online) Particle population for the hadron (a) and quark (b) phases, and the corresponding matter magnetisation, as a function of the baryon chemical potential for a constant MF, $\sim 5 \times 10^{18} \, \mathrm{G}$. In panel~(a), it can be seen how the peaks and the valleys of the magnetisation (grey curve) are related to the appearance or disappearance of hadron particles.  In panel~(b), the magnetisation responds to the oscillation of electron population which follows the Landau level occupation. To improve the view of the magnetisation effect, we have painted a cyan area below the magnetisation grey curve.}
    \label{fig:magabu}
\end{center}
\end{figure}

\begin{figure}
\begin{center}
	\includegraphics[width=0.80\columnwidth]{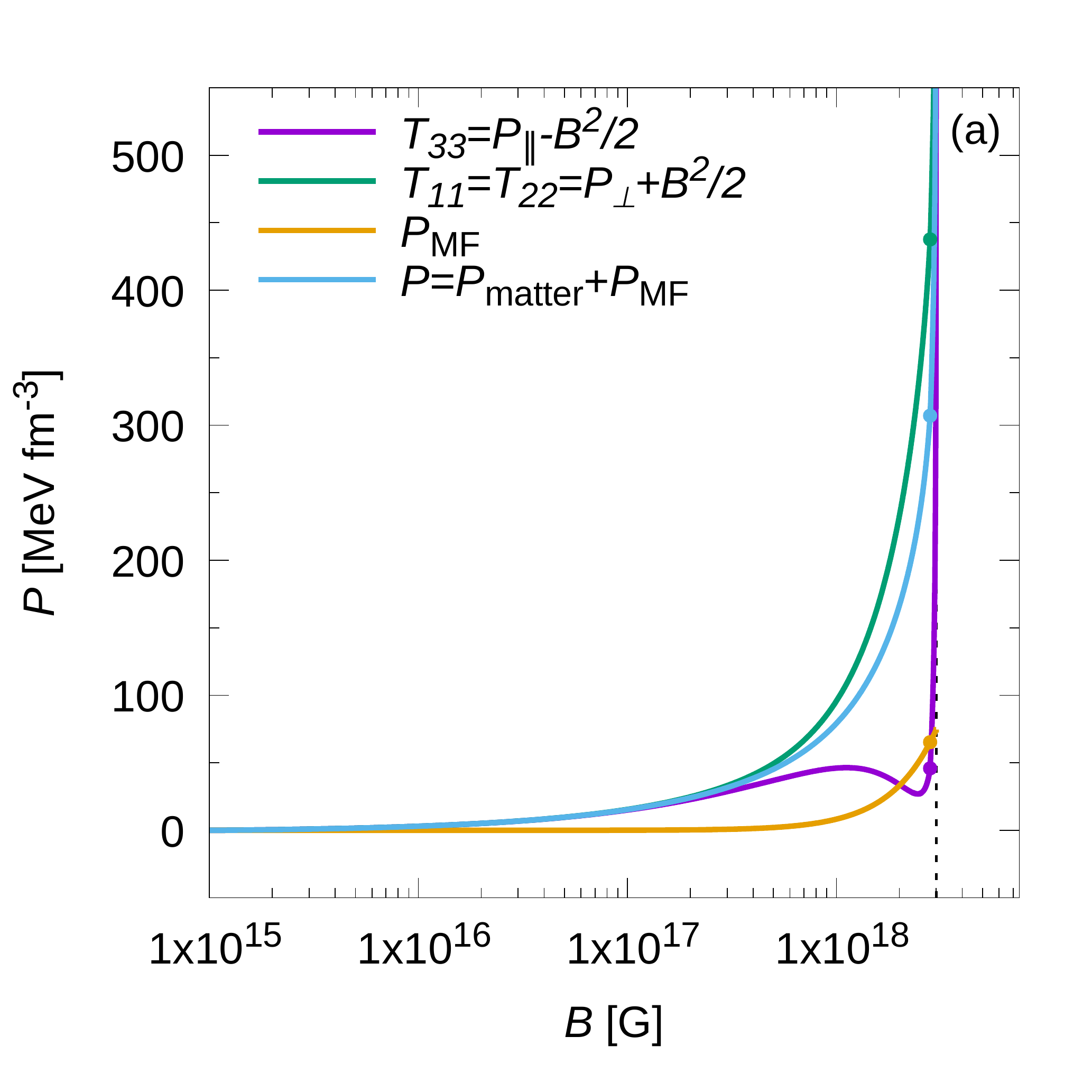}\\
	\includegraphics[width=0.80\columnwidth]{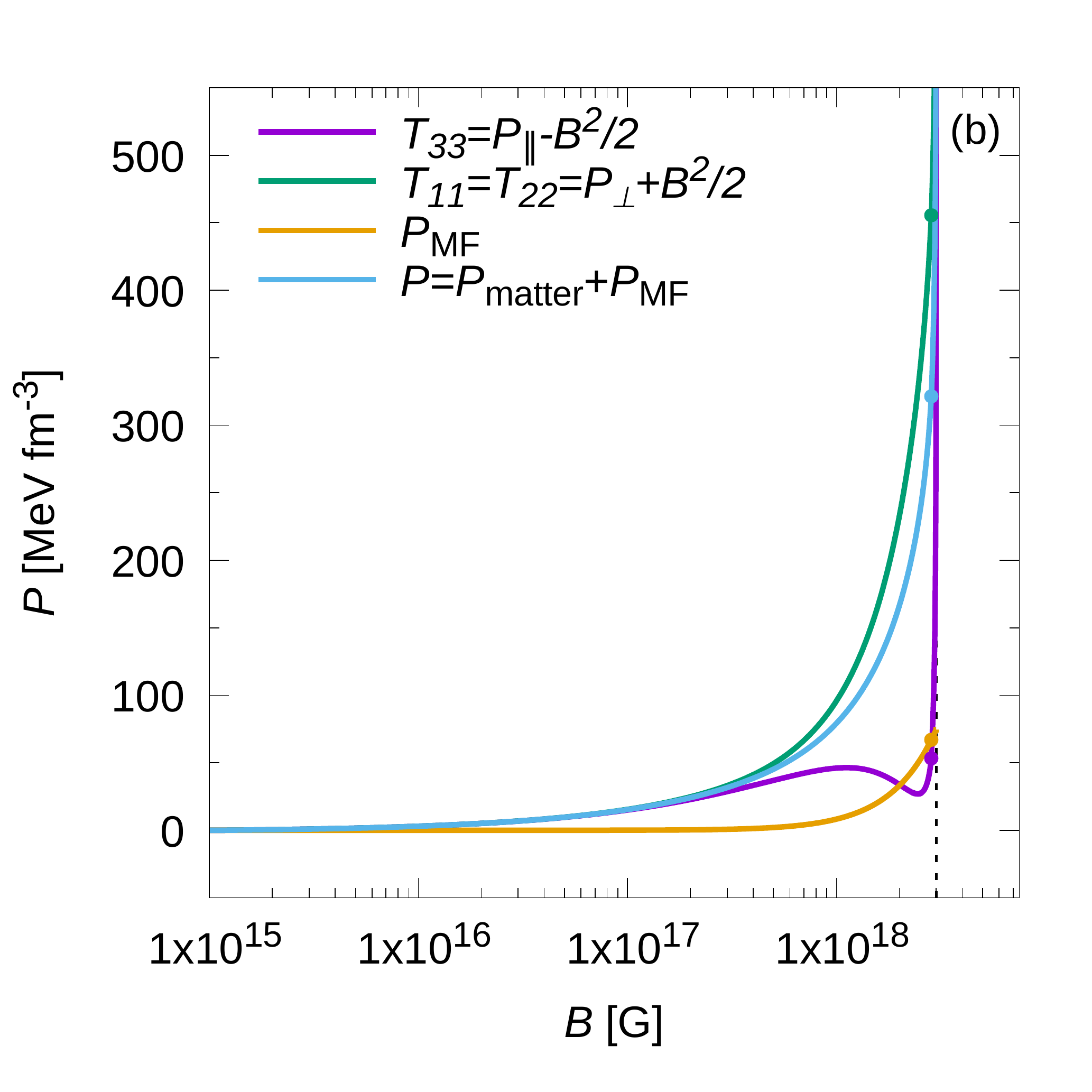}
    \caption{(Colour online) Pressure components for hybrid matter as a function of the MF for Set 1  (a) and Set 2 (b) of the FCM parameters and for the magnetar case. The dot in each curve marks the phase transition  from hadron to quarks, so that for low densities there is hadron matter and for high densities, quark matter. It can be seen that the parallel and transverse pressures are equal for low MF and they split increasingly as the MF increases. It also can be seen that the pure magnetic pressure dominates only in the extremely high MF regime. The low-B HS case is not shown because it has no anysotropies in the pressure and it has a negligible contribution of the magnetic field pressure term for the whole MF domain.}
    \label{fig:pppt1}
\end{center}    
\end{figure}

Local conservation of baryon and electric charges are given by
\begin{align}
& \sum_i q_{b,i} n_i = n_B \, , \nonumber \\
& \sum_i q_{e,i} n_i + \sum_l q_{e,l} n_l = 0 \, ,
\end{align}
respectively, where the sum of the first equation and the first sum of the second equation go over the baryons or quarks, depending on the phase, and the second sum of the second equation goes over the leptons, $n_i$ are the respective number densities, $q_{b,i}$ and $q_{e,i}$ are the baryon and electric charges of the respective baryon and $n_B$ is the total baryon number density.

\begin{figure}
\begin{center}
	\includegraphics[width=0.8\columnwidth]{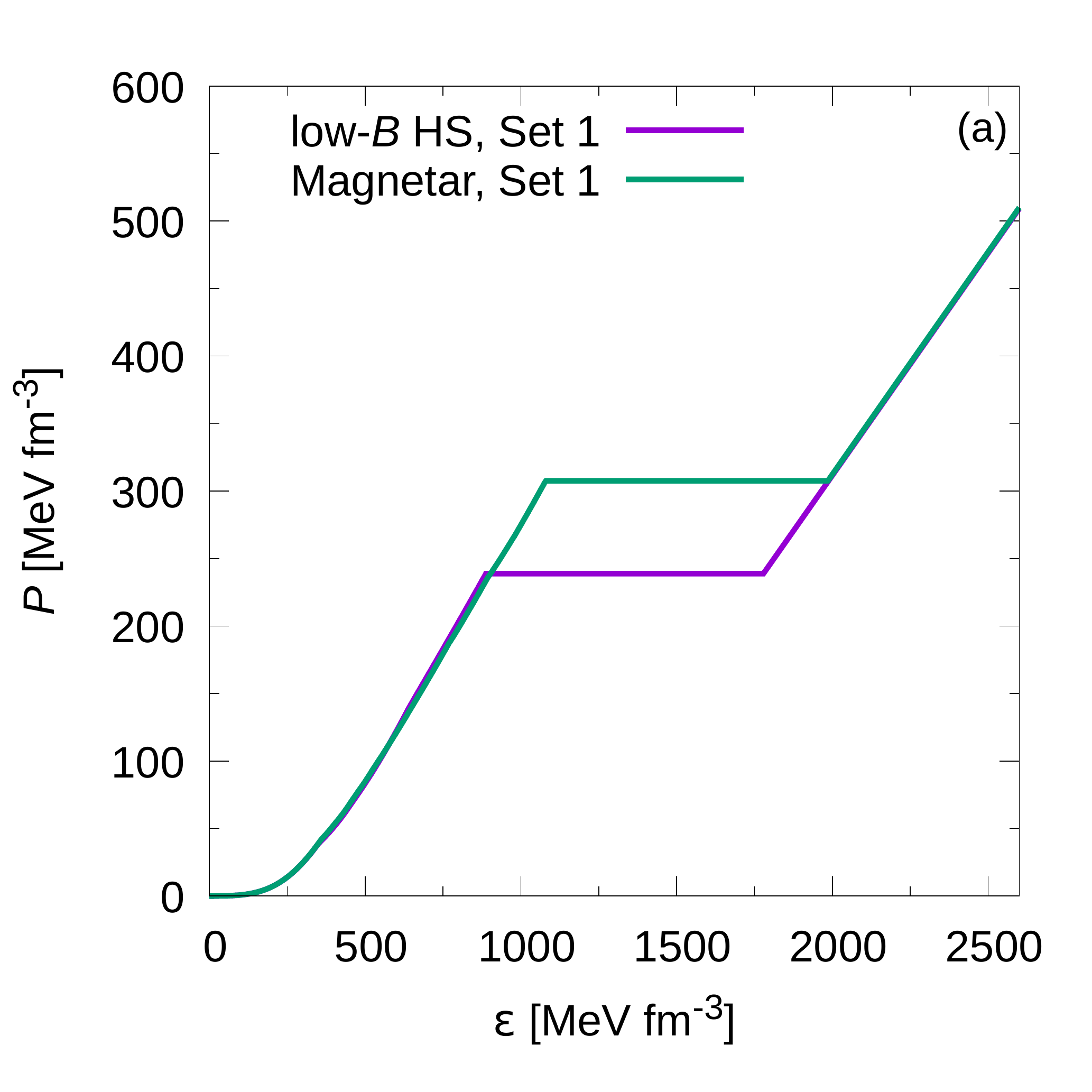}
	\includegraphics[width=0.8\columnwidth]{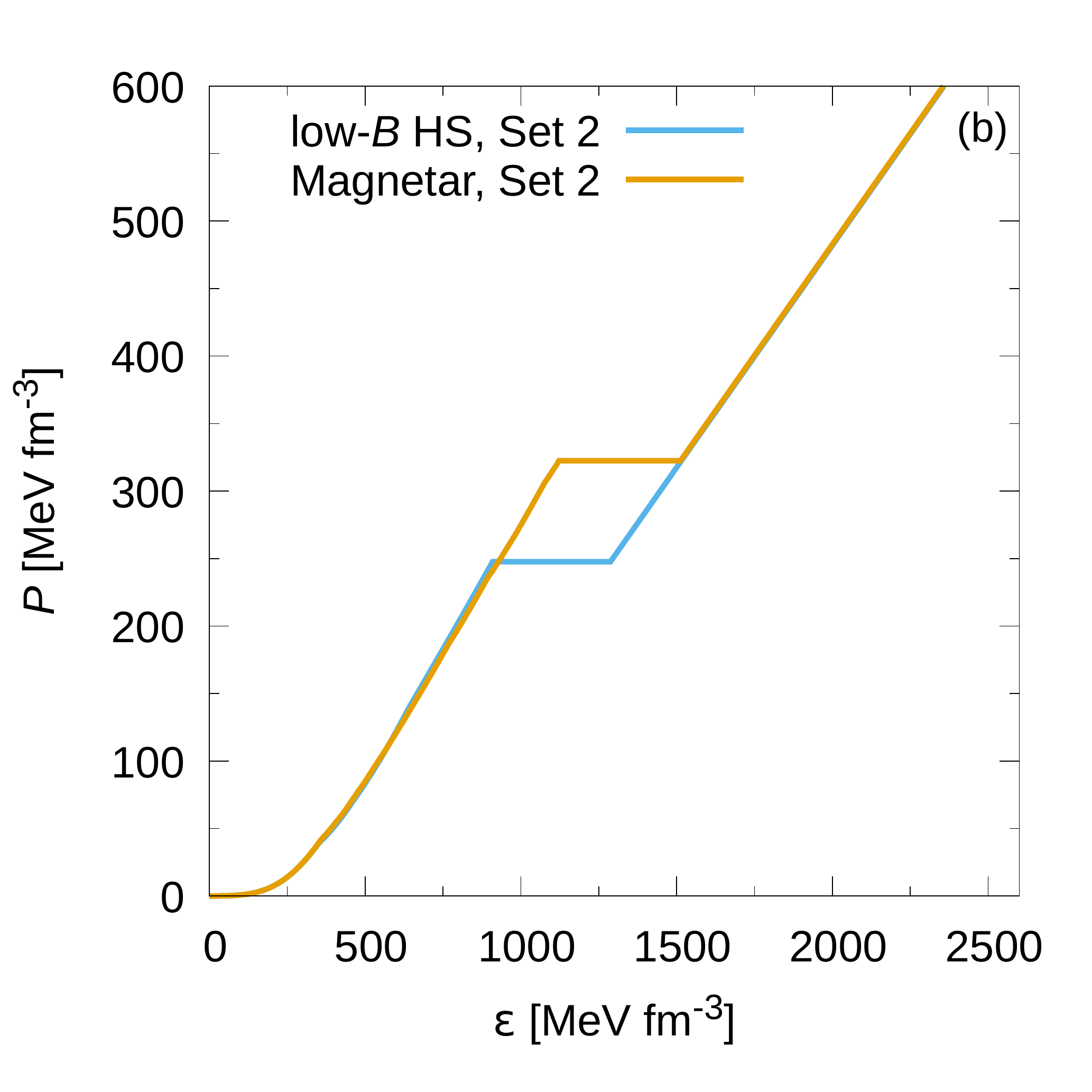}
    \caption{(Colour online) Hybrid EoSs for the two cases, \mbox{low-$B$ HS} and magnetar, and for Set 1 (a) and Set 2 (b) of the FCM parameters. In the curves, the constant pressure regions correspond to the phase transition.}
    \label{fig:eos}
\end{center}
\end{figure}

In  $\beta$-equilibrium, the relationships among the chemical potentials is given by
\begin{equation}
\mu_i = q_{b,i} \, \mu_B - q_{e,i} \, \mu_e \, ,
\end{equation}
where $\mu_i$ is the chemical potential of each particle (baryons, quarks or leptons), $\mu_B$ is the baryon chemical potential and $\mu_e$ is the electronic chemical potential. Under these chemical equilibrium conditions, the thermodynamic quantities of each phase can be calculated and the hybrid EoS constructed.

The first aspect we shall discuss about the EoS results is the contribution of the magnetisation, $\mathcal{M}$, and the magnetisation pressure, $- \mathcal{M} B$.  Fig.~\ref{fig:premag} shows the magnetisation pressure for both phases, hadron, panel~(a), and quark, panel~(b), as a function of the baryon chemical potential and the magnetic field. We remark that, for this particular calculation, we are not considering the parametrisation of Eq.~(\ref{param}), but we take the magnetic field and the baryon chemical potential as independent variables. In addition, it is important to point out that for these results we are not considering the FCM parameters of Table \ref{tab3}. To show and analyse  the effects of the magnetisation, we take $V_1 = 20$~MeV and $G_2= 0.006$~GeV$^4$, which are representative of the general behaviour of the quark model.

Let us analyse the magnetisation of both, the hadron and the quark phases separately. Fig.~\mbox{\ref{fig:premag}(a)} shows that the hadron magnetisation pressure is significant only for extremely high values of the MF, in a density regime where hadrons are no longer present because of the occurrence of the phase transition to pure quark matter. Because of this behaviour and to avoid an unnecessary  complication of the hadron EoS, we only consider the magnetisation pressure term in the quark phase. It is worth to mention that for the quark phase in Fig.~\mbox{\ref{fig:premag}~(b)}, the magnetisation and the magnetisation pressure can be calculated analytically from the pressure expressions, $\mathcal{M} B=P_{\parallel}-P_{\perp}$. However, to obtain the same quantity for the hadron phase, we have to numerically derive the magnetisation from Eq.~(\ref{ppareq}).

The results shown in Fig.~\ref{fig:magabu} deepen our study of magnetised matter. There, it can be seen the magnetisation and the particle populations as a function of the baryon chemical potential for a given constant MF for hadron matter in panel~(a) and quark matter in panel~(b). It is noticeable how the magnetisation reacts to the changes in the particle populations. In the hadron phase, the peaks and the valleys of the magnetisation occur due to the appearance or the vanishing of each type of particle. In the quark phase, the magnetisation has an oscillatory behaviour correlated to the oscillations of the particle populations, more evident for electrons, indicating the occupation of new Landau levels with the increase of the chemical potential.

In Fig.~\ref{fig:pppt1}, we show the different components of the pressure for hybrid matter as a function of the MF for the magnetar EoSs and two sets of the FCM parameters given in Table \ref{tab3}. It can be seen that both pressure contributions differ  from each other as the MF increases. The pure magnetic pressure becomes relevant only for high MF.
The \mbox{low-$B$ HS} case is not shown because there are no anisotropies in the pressure components and the pure magnetic field pressure term is negligible for the whole MF domain. 

Finally, we present results for  magnetised hybrid matter. In Fig.~\ref{fig:eos} we present the magnetised hybrid EoS for \mbox{low-$B$ HSs} and magnetars and for the two sets of FCM parameters. It can be seen how the variation of the MF or the FCM parameters affects the transition pressure and the energy gap between the hadron and quark phases. Comparing panels (a) and (b) it can also be noted that the jump in the energy density at the hadron-quark phase transition for Set 1 is bigger than for Set 2. For both sets of FCM parameters, when the magnetic field is stronger, as in the case of magnetars, the transition pressure is higher and occurs at higher densities. 

\begin{figure}
\begin{center}
	\includegraphics[width=0.95\columnwidth]{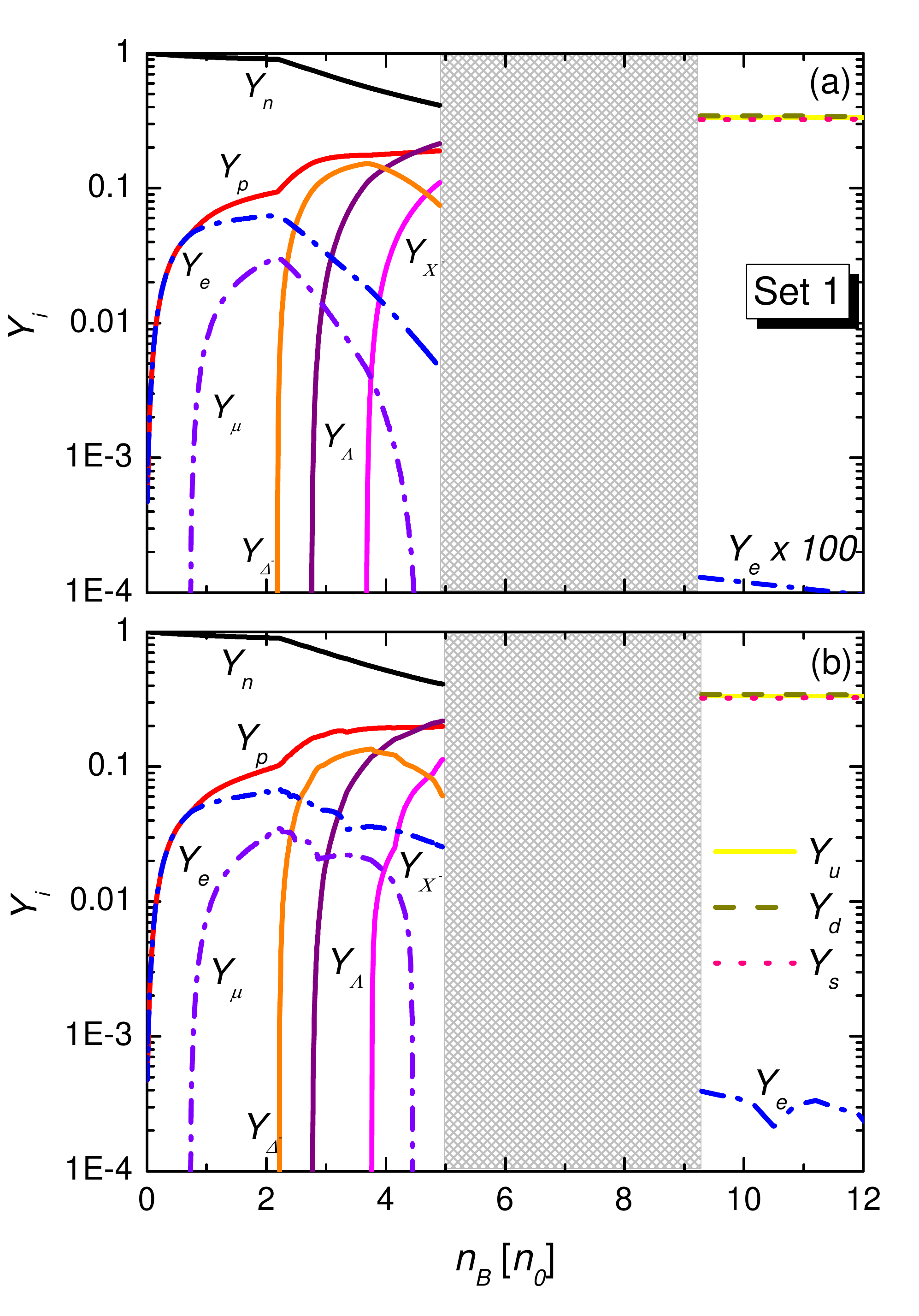}
    \caption{(Colour online) Particle population, $Y_i$, as a function of the baryon number density in units of nuclear saturation density, $n_0$, for \mbox{low-$B$ HSs} (a) and magnetars (b). The parameters considered for the quark phase of the hybrid EoS are $V_1 = 20$~MeV and $G_2 = 0.014$~GeV$^4$ (Set 1). The grey area indicates the jump in the density due to the phase transition.}
    \label{fig:abu1}
\end{center}    
\end{figure}

\begin{figure}
\begin{center}
	\includegraphics[width=0.95\columnwidth]{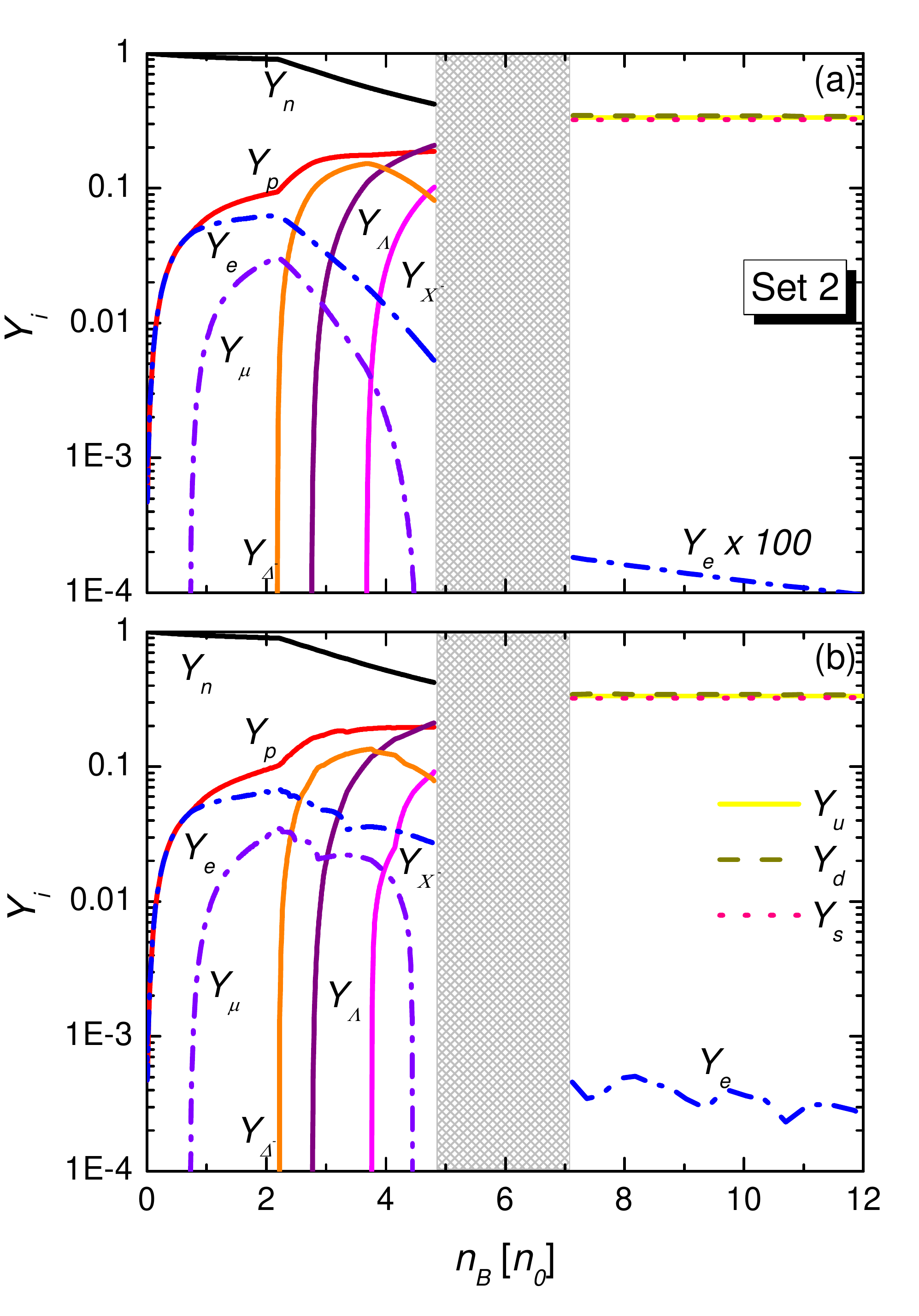}
    \caption{(Colour online) Same as in the previous Figure but for $V_1 = 90$~MeV and $G_2 = 0.007$~GeV$^4$ (Set 2).}
    \label{fig:abu2}
\end{center}    
\end{figure}

In Figs.~\ref{fig:abu1} and \ref{fig:abu2} we show the resulting hybrid particle populations as a function of the baryon number density for \mbox{low-$B$ HSs} and magnetars and for the two sets of FCM parameters. As can be seen, the jump in the number density at the transition (the grey area) is also very dependent of the FCM parameters values  but the MF strength effect is not as noticeable as in  Fig.~\ref{fig:eos}. Although the variation of the MF strength or the FCM parameters affect the amount of particle abundances, they do not modify the kind of particles appearing in the hadron phase before the phase transition. Also, the $\Delta^-$ resonances have a strong presence in all the cases, appearing before most of the particles of the baryon octet. On the other hand, as usual, there is a strong drop of the lepton abundances from the hadron phase to the quark phase. However, for magnetars, this drop is smaller than for low-B HSs.

\section{Stellar structure and stability}
\label{structure}
\subsection{Hydrostatic equilibrium configurations}

Once we construct the hybrid magnetised EoS, we can obtain several results for our model solving the equilibrium structure equations for the HS. As we have already stated and in order to obtain representative, suitable and distinguishable star configurations, we select two cases for the MF parametrisation, a \mbox{low-$B$ HS}, with $B_{min} = 10^{13}$~G and $B_{max} = 10^{15}$~G, and a magnetar, with $B_{min} = 10^{15}$~G and $B_{max} = 3 \times 10^{18}$~G. Besides, the chosen FCM parameters sets of Table \ref{tab3} fulfil the $2~M_\odot$ constraint.

\begin{table} 
	\centering
	\caption{Sets of FCM parameters chosen to calculate the stellar models.}
	\begin{tabular}{cccc} 
		\hline
		Set & $V_1$ {[MeV]} & $G_2$ [GeV$^4$] \\
		\hline
		1 & 20 & 0.014 \\
		2 & 90 & 0.007 \\
		\hline
	\end{tabular} \label{tab3}
\end{table}

According to the explanation in Section~\ref{sec:average}, to construct stable hydrostatic stellar configurations, we can use the TOV equations as a valid approximation, which are given by
\begin{equation}
\label{tov1}
\frac{dP}{dr} = -\frac{G m(r) \epsilon(r)}{r^2} \frac{ [1 + P(r) / \epsilon(r) ] [1 + 4 \pi r^3 P(r) / m(r)]}{1-2 G m(r)/r} \,,
\end{equation}
\begin{equation}
\label{tov2}
\frac{dm}{dr} = 4 \pi r^{2} \, \epsilon(r) \,,	
\end{equation}
where $r$ is the radius, $P(r)$ and $\epsilon(r)$ are the pressure and energy density at a radius $r$, $m(r)$ is the mass bounded by the radius $r$ and $G$ is the universal gravitational constant. The conditions required to solve the integrals are:
\begin{equation}
	P(r=R)=0\, \text{,} \, m(r=0)=0 \, .
\end{equation}

Hence, using the magnetised hybrid EoS as an input, we can obtain the mass-radius and mass-central energy density relationships for a family of stars, each one possessing a different central energy density for the integration procedure.

\subsection{Dynamical stability of hybrid stars with slow and rapid phase transitions at the quark-hadron interface}
\label{sec:dynamical_stability}

The stability of a star in hydrostatic equilibrium can be assessed by means of the analysis of its response to small radial perturbations \citep{Chandrasekhar}. When a \textit{stable} star is perturbed, fluid elements all along the stellar interior oscillate around their equilibrium positions, compressing and expanding periodically. On the contrary, small perturbations grow without limit in the case of an \textit{unstable} star, leading to its collapse or disruption.

In the case of hybrid stars, special care must be taken with fluid elements in the neighbourhood of the quark-hadron interface,  because, as the fluid oscillates, their pressures become alternatively higher and lower than the phase transition pressure $P_t$. The quark-hadron transition is a quite complex mechanism involving strong interactions, surface and curvature effects, Coulomb screening, etc. (see \cite{Lugones2} and references therein); thus, a fluid element oscillating around $P_t$ will not necessarily undergo a phase conversion. The probability that such conversion occurs depends on the nucleation timescale, which at present is a model dependent quantity with an uncertain value (see \citet{Lugones2,Bombaci2,Lugones1} and references therein). In order to analyse the stability of hybrid stars,  we will focus here on two limiting cases: \textit{slow} and \textit{rapid} phase transitions. Slow conversions at the interface involve the stretch and squash of volume elements near the quark-hadron interface without their change of nature (nucleation timescales much larger than those of perturbations). Rapid conversions imply a practically immediate conversion of volume elements from one phase to the other and vice-versa in the vicinity of the discontinuity upon any perturbation \citep*{Haensel,Pereira1}. 

In spite of being physically complex, the nature of the conversion can be mathematically summarised into simple \textit{junction conditions} on the radial fluid displacement $\xi$ and  the corresponding Lagrangian perturbation of the pressure $\Delta P$ at the phase-splitting surface \citep*{Pereira1}. For slow conversions, the jump of $\xi$ and $\Delta P$ across the interface should always be null:
\begin{equation}
[\xi_s ]^+_{-}\equiv \xi_{s}^+-\xi_{s}^- =0 \qquad [\Delta P]^+_- \equiv \Delta P^+ - \Delta P^- =0 .
\label{xislow}
\end{equation}
For rapid phase transitions we have 
\begin{equation}
[\xi_r]^+_-=\Delta P\left [\frac{1}{P'_0}\right]^+_-   \qquad \qquad [\Delta P]^+_- =0  ,
\label{xirapid}
\end{equation}
where $P_0' \equiv dP_0/dr$ is the pressure gradient of the background pressure at the interface. 

An analysis of the radial oscillation spectrum of a spherically symmetric star shows that, if the fundamental frequency is a real number ($\omega_{0}^{2}>0$), then any radial perturbation of the star will produce oscillatory fluid displacements and the stellar configuration is dynamically stable \citep{Chandrasekhar}. Fortunately, in many cases, a much simpler static condition can be derived from the latter one. In fact, it is widely known that a hydrostatic configuration of a spherically symmetric cold-catalysed one-phase star is unstable if the condition 
\begin{equation}
\partial M/\partial \epsilon_c < 0   
\label{dMdE_c}
\end{equation}
is verified, where $\epsilon_c$ is the central density of a star whose total mass is $M$ \citep{Harrison}. Notwithstanding, this is not necessarily the case if matter is non-catalysed \citep{Gourgoulhon} or if the system contains multiple phases separated by sharp density discontinuities \citep{Haensel,Pereira1,Pereira2}.  In particular, it has been shown in \citet*{Pereira1},  that the standard stability criterion of Eq.~\eqref{dMdE_c} remains always true for rapid phase transitions but breaks down in general for slow phase transitions. In fact, for slow transitions the frequency of the fundamental mode can be a real number (indicating stability) even for some branches of stellar models that verify $\partial M/\partial \epsilon_c < 0$.  Thus, in the case of slow conversions, new branches of stable stellar configurations could arise.

\subsection{Microphysics of phase conversions}

\begin{figure}
\begin{center}
	\includegraphics[width=0.90\columnwidth]{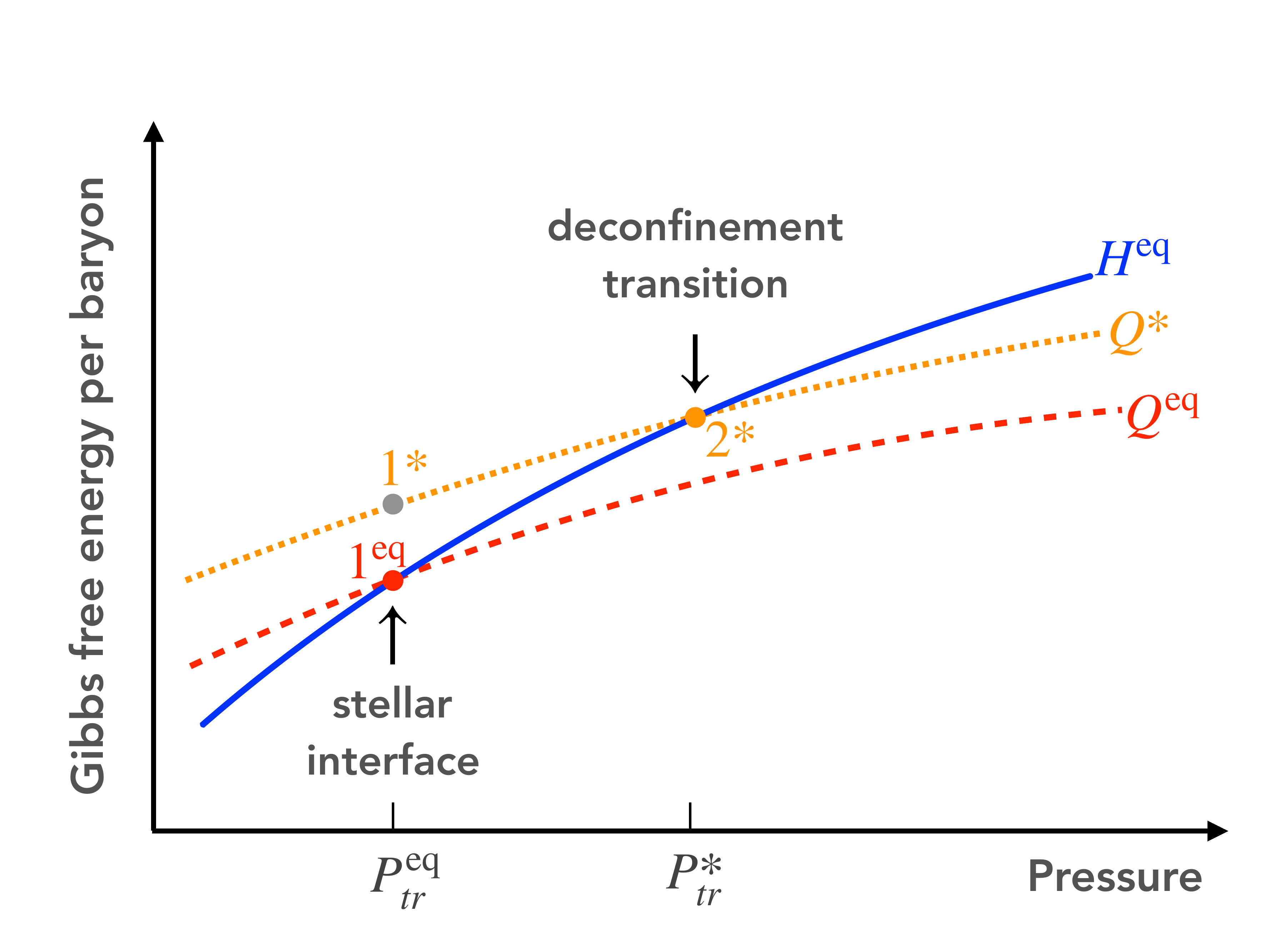}
    \caption{(Colour online)  Schematic representation of the Gibbs free energy per baryon of hadron and quark matter in chemical equilibrium under weak interactions ($H^{\mathrm{eq}}$ and $Q^{\mathrm{eq}}$ respectively), and just deconfined quark matter out of chemical equilibrium ($Q^{*}$). The phase $Q^{*}$ has the same flavor composition than $H^{\mathrm{eq}}$ at the same pressure (see \citet{Olesen:1993ek, Lugones:1997gg, Iida:1998pi, Bombaci:2004mt, Lugones:2011xv}).  The conversion process is rapid if pressures $P_{tr}^{\mathrm{eq}}$ and $P_{tr}^{*}$ are very close to each other. If not, the conversion would be slow. }
\label{fig:gibbs}
\end{center}    
\end{figure}

In the following, we analyse the hadron-quark conversion of perturbed fluid elements around the hybrid star's interface from a microscopic point of view. The interface is  at some pressure $P_{tr}^{\mathrm{eq}}$ at which the Gibbs free energy per baryon of hadron matter  ($H^{\mathrm{eq}}$) and quark matter  ($Q^{\mathrm{eq}}$) are equal (see Fig. \ref{fig:gibbs}). Both phases ($H^{\mathrm{eq}}$ and $Q^{\mathrm{eq}}$) are individually in chemical equilibrium under weak interactions. For pressures below $P_{tr}^{\mathrm{eq}}$ the $H^{\mathrm{eq}}$ phase is favoured because of its lower free energy and for pressures above $P_{tr}^{\mathrm{eq}}$ the $Q^{\mathrm{eq}}$ phase is preferred. Let us consider a small lump of hadron matter in the neighbourhood of the interface, resting in equilibrium at the ``hadronic side'' of the discontinuity. If the star is unperturbed, this lump is in thermal, mechanical and chemical equilibrium, as any other fluid element inside the star. In particular,  chemical equilibrium under weak interactions determines univocally the particle abundances of all baryon and lepton species at that place (see Figs. \ref{fig:abu1} and \ref{fig:abu2}). When this small hadronic lump is perturbed radially, it moves to the ``quark side'' of the interface, which is at a pressure slightly above $P_{tr}^{\mathrm{eq}}$ (see Fig. \ref{fig:gibbs}).  At this new position, it is submitted to an overpressure that could convert hadrons into quarks if the deconfined quark-matter state had a lower Gibbs free energy per baryon. However, since deconfinement is driven by strong interactions, the quark and lepton abundances $Y_i$ of \textit{just} deconfined quark matter are determined by the composition of the hadronic phase from which the new state emerges; more specifically, by the flavour conservation conditions given in e.g. \citet{Olesen:1993ek} and  \citet{Lugones:1997gg}. As a consequence, just deconfined matter would be out of chemical equilibrium,  and cannot be located along the $Q^{\mathrm{eq}}$ curve in Fig. \ref{fig:gibbs}. Model calculations \citep{Olesen:1993ek, Lugones:1997gg, Iida:1998pi, Bombaci:2004mt, Lugones:2011xv} show that, if quark matter is to be formed under flavour conservation conditions around the pressure $P_{tr}^{\mathrm{eq}}$,  then its state would be represented by a point $1^{*}$ on the curve $Q^*$,  which is always above the point $1^{\mathrm{eq}}$ of phase coexistence in chemical equilibrium  (see Fig. \ref{fig:gibbs}). 
Since the transition to the point $1^{*}$ is energetically suppressed, the perturbed hadronic lump will move to the right along the solid blue line of Fig. \ref{fig:gibbs}. A phase conversion would be possible if the lump is able to reach the point $2^{*}$ where the free energy of $H^{\mathrm{eq}}$ and $Q^{*}$ are equal. Therefore, the phase conversion would be rapid if the pressures $P_{tr}^{\mathrm{eq}}$ and $P_{tr}^{*}$ are close enough to each other and slow if they are sufficiently different. Since first principle calculations of the high-density EoS are still unavailable, a conclusive answer to this issue is missing. However, numerical results using different phenomenological models \citep{Iida:1998pi, Bombaci:2004mt, Lugones:2011xv} show that the relevant nucleation times are huge, strongly suggesting that the hadron-quark conversion near the interface is slow. A reasoning similar to the one given before holds for the conversion of a quark lump into hadron matter at the hybrid star interface.

\subsection{Slow phase conversions: terminal configuration and determination of new stable stellar branches}

In order to determine the existence of new stable stellar branches in the case of slow conversions, it would be necessary to calculate the fundamental radial oscillation mode for each equilibrium configuration. However,  since unstable stars are characterised by purely imaginary frequencies ($\omega_0^2<0$),  what is important for stability purposes is to focus on the case $\omega_0^2=0$. To this end, we set $\omega=0$ in the radial oscillation equations. The result (adopting $G = c = 1$) is: 
\begin{align} 
\frac{d \xi}{d r} &=V(r) \xi+W(r) \Delta P , \label{oscillation1}\\ \frac{d \Delta P}{d r} &=X(r) \xi+Y(r) \Delta P , \label{oscillation2}
\end{align}
where the coefficients are given by
\begin{align} 
V(r) &= -\frac{3}{r}-\frac{dP}{dr}\frac{1}{(P+\epsilon)},  \\
W(r) &= -\frac{1}{r}\frac{1}{\Gamma P}   \\
X(r) &= -4\frac{dP}{dr} + \bigg(\ \frac{dP}{dr}\bigg)^{2}\frac{r}{(P+\epsilon)}-8\pi \mathrm{e}^{\lambda}(P+\epsilon) Pr    \\
Y(r) &= \frac{dP}{dr}\frac{1}{(P+\epsilon)}- 4\pi(P+\epsilon)r \mathrm{e}^{\lambda} \label{ecuacionparaP}.
\end{align}

From the previous equations we can determine the properties (mass, radius, central energy density, etc.) of the star for which the frequency of the radial fundamental mode is null. Since now such stellar configuration is not necessarily the one with the maximum mass, we shall call it \textit{terminal configuration}, and its mass will be denominated \textit{terminal mass},  $M_T \equiv M(\omega^2_0=0)$. To determine $M_T$ we solve Eqs.~\eqref{oscillation1}--\eqref{oscillation2} simultaneously with the TOV equations (\ref{tov1})--(\ref{tov2}) for a given central energy density and use the initial conditions $\xi(0)= \textrm{constant} $ and $\Delta P(0)= -3\Gamma(0) P(0)\xi(0)$ for the (now auxiliary variables)  $\xi$ and  $\Delta P$. In this procedure one has not to worry about the conditions $[\xi_s]^+_-=0$ and $[\Delta P]^+_-=0$ at the phase-splitting surface because they are automatically fulfilled along the numerical integration (with a piecewise EoS satisfying Gibbs criteria for phases in equilibrium). When arriving at the stellar surface, $\Delta P$ will not be zero in general, because we are not solving now a Sturm-Liouville problem. Spanning the value of the central density, one will arrive at a stellar configuration for which the condition $\Delta P =0$ is verified at the star's surface. Such configuration is a terminal one if and only if the corresponding function $\xi(r)$ has no nodes inside the star. If $\xi (r)$ has any node, it means that the stability of an excited mode is changing (stabilising or destabilising)  for that configuration. 

\begin{figure}
\begin{center}
	\includegraphics[width=0.85\columnwidth]{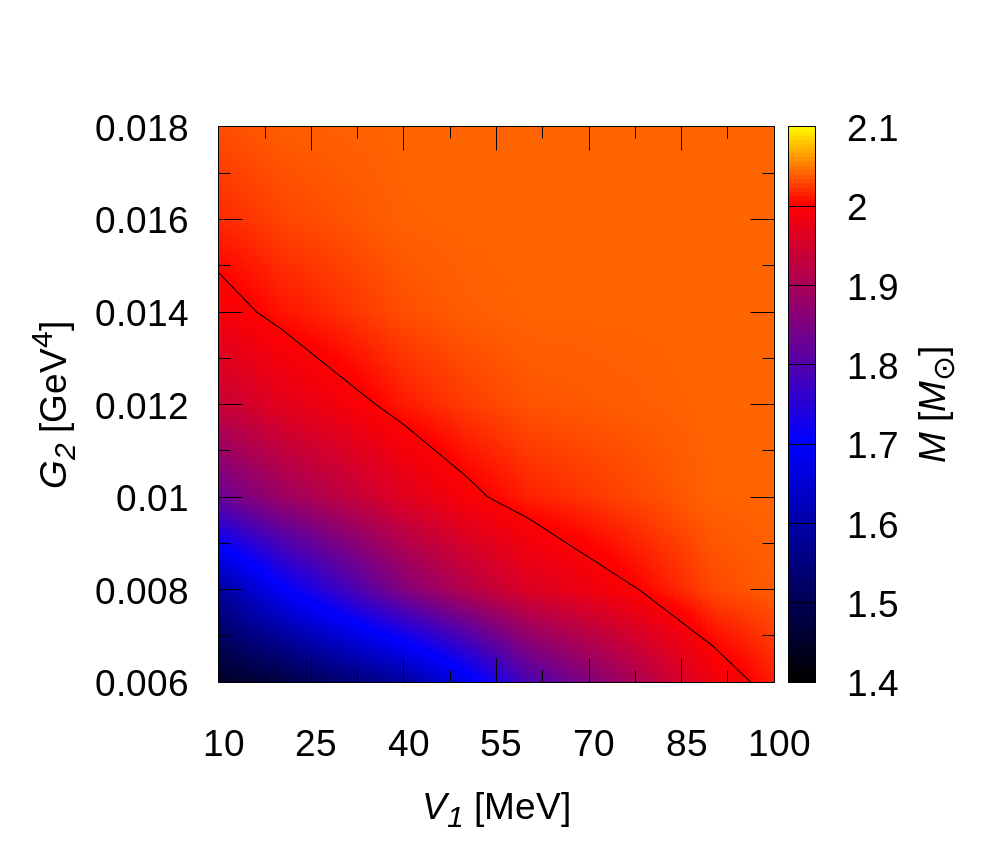}
    \caption{(Colour online) Maximum mass for each family of stars as a function of $V_1$ and $G_2$. This result is obtained for \mbox{low-$B$ HSs} in order to determine the FCM parameters values that fulfil the pulsar observations constrain. The curve marks the contour line for $M_{\rm{max}} = 2~M_\odot$.}
    \label{fig:mmax1}
\end{center}    
\end{figure}

\subsection{Results: equilibrium configurations and dynamical stability}

In Fig.~\ref{fig:mmax1} we show results of solving the TOV equations for a wide range of the $V_1$ and $G_2$ FCM parameters and we plot the maximum mass in the $V_1-G_2$ plane for the \mbox{low-$B$ HS} model. The black line represents the set of parameters for which the maximum mass is $2~M_\odot$. In the left bottom corner of Fig.~\ref{fig:mmax1} we see that the maximum mass increases with both parameters,  but then it tends to a constant value as one approaches the top right corner of of Fig.~\ref{fig:mmax1}.

\begin{figure}
\begin{center}
	\includegraphics[width=0.85\columnwidth]{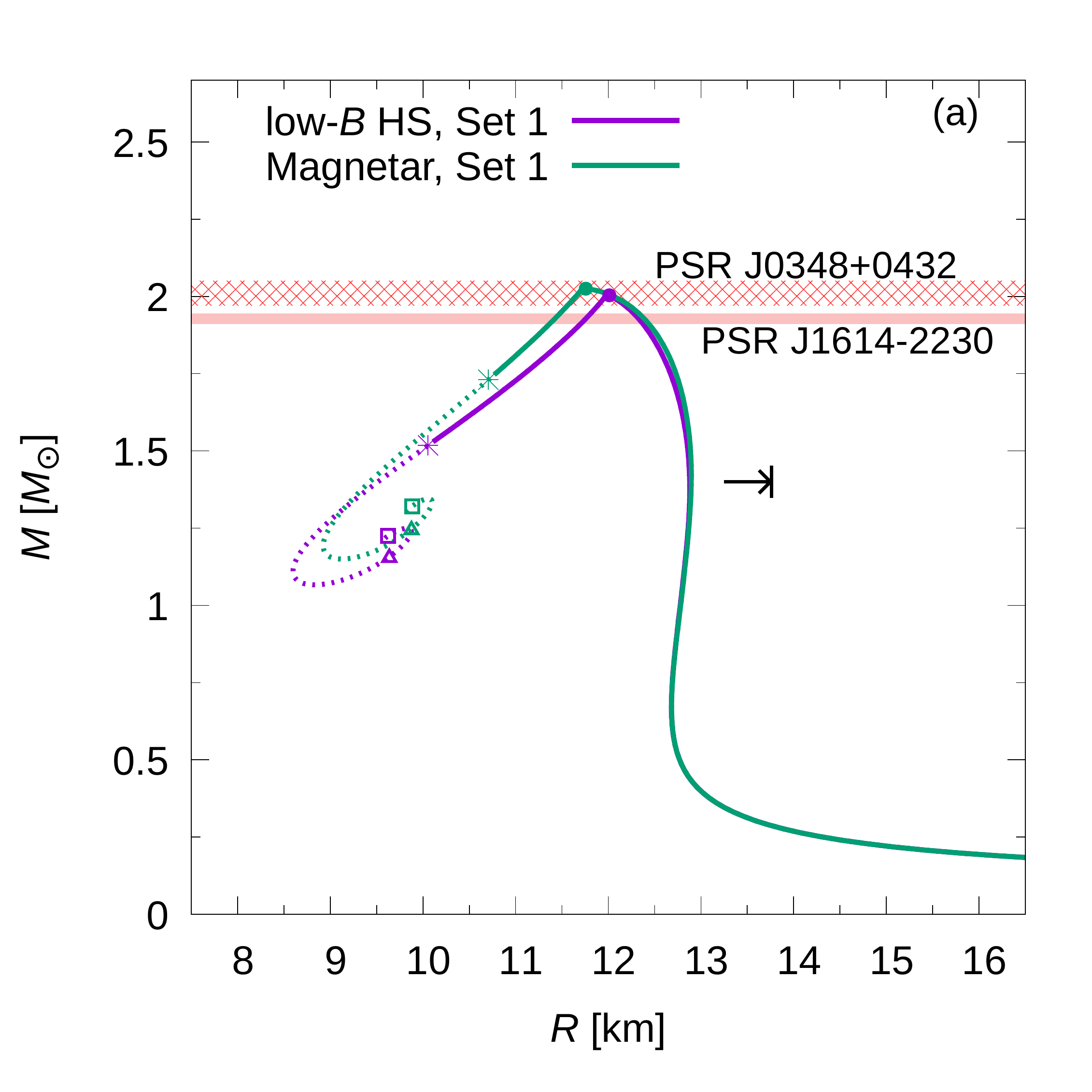}\\
	\includegraphics[width=0.85\columnwidth]{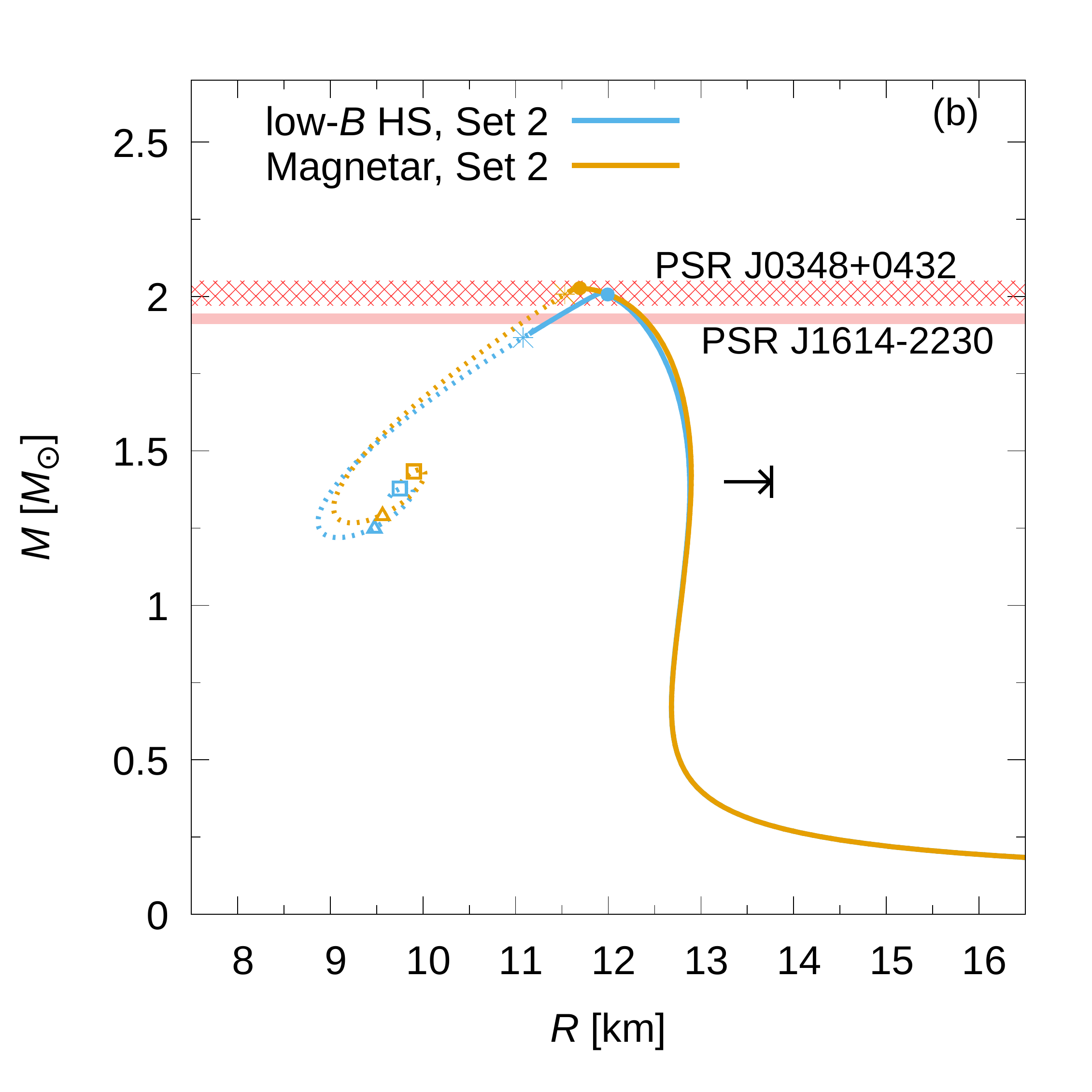}
    \caption{(Colour online) $M$--$R$ relationship for \mbox{low-$B$ HSs} and magnetars, for Set 1 (a) and Set 2 (b) of the FCM parameters. The round dot indicates the place where the quark matter core appears. In the case of rapid transitions, all models to the left of the maximum mass are unstable. In the case of slow transitions, the dashed branch to the left of the asterisk (terminal configuration) represents the unstable models because $\omega_0^2<0$ there.  The triangle dot marks where the first radial mode becomes unstable and the square dot marks where the second radial mode becomes unstable. The horizontal bars are the measured masses of the $2~M_\odot$ pulsars with their corresponding errors \citep{Antoniadis,Arzoumanian_2018}. The black horizontal arrow marks the constrain calculated by \protect\cite{Fattoyev} for GW170817, $R_{1.4} < 13.76 \, \mathrm{km}$.}
    \label{fig:mraio}
\end{center}    
\end{figure}

\begin{figure}
\begin{center}
\includegraphics[width=1.00\columnwidth]{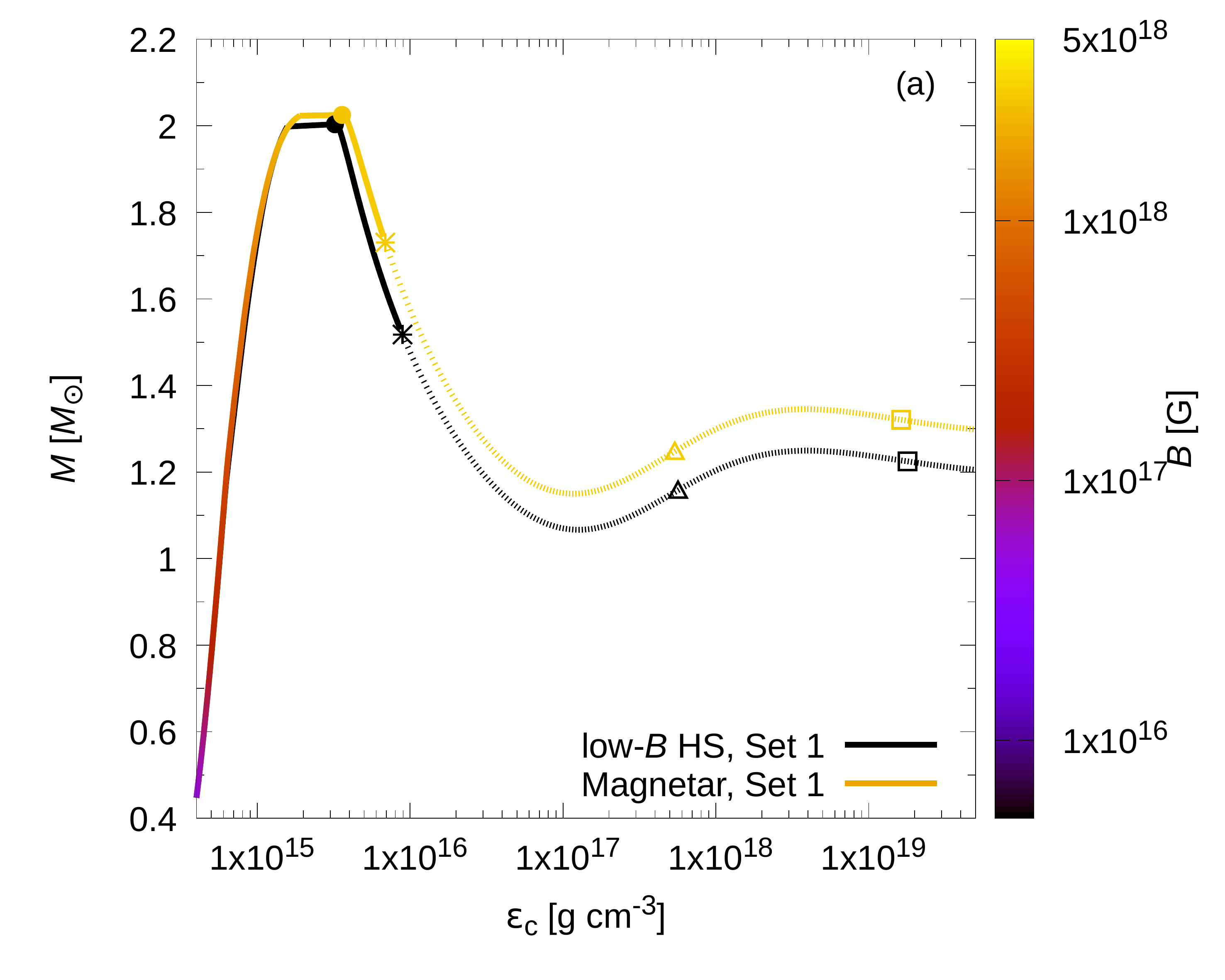}
\includegraphics[width=1.00\columnwidth]{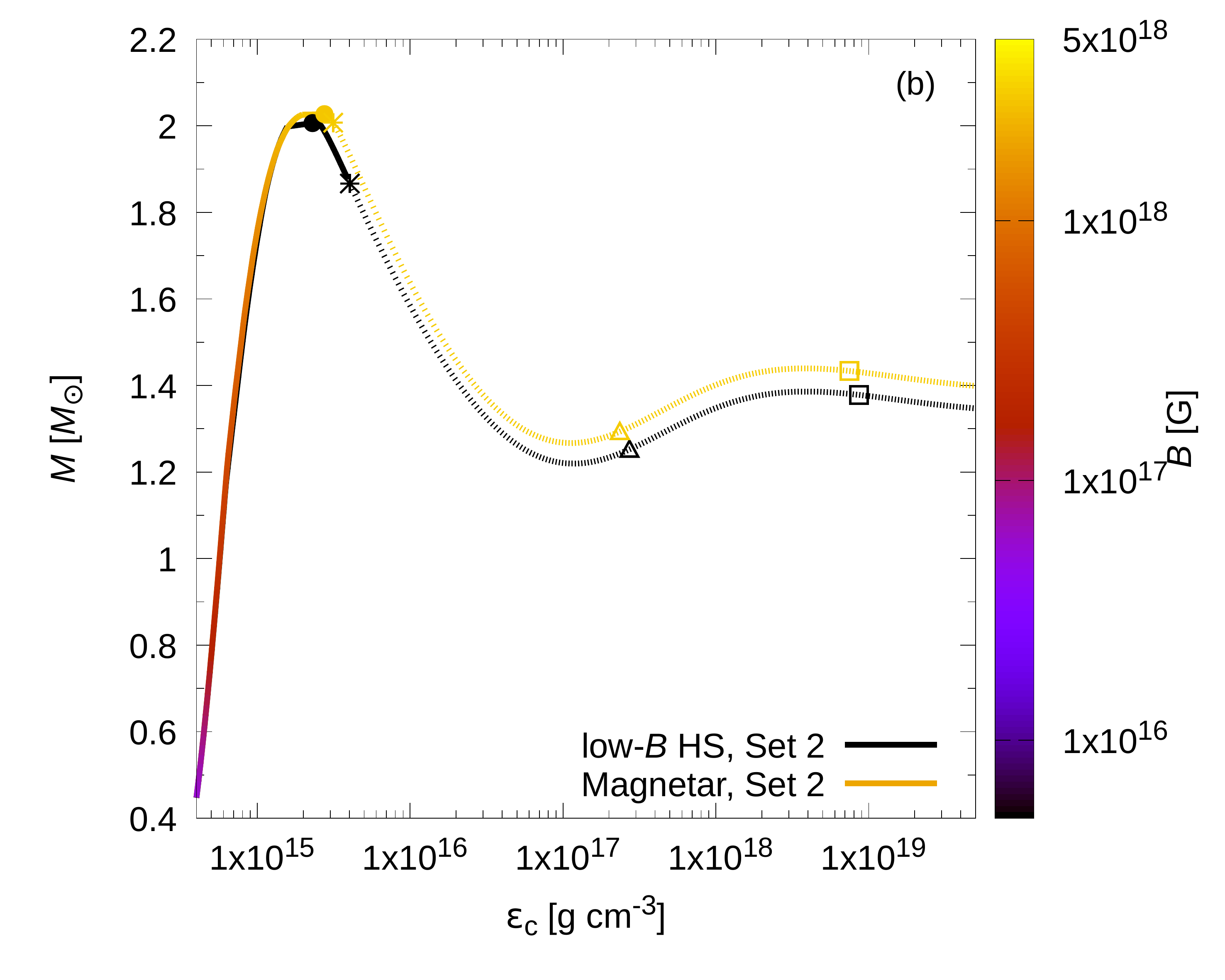}
    \caption{(Colour online) Gravitational mass $M$ versus central energy density $\epsilon_c$ relationship for  \mbox{low-$B$ HSs} and magnetars, for Set 1 (a) and Set 2 (b) of the FCM parameters. The colour scale indicates the value of the magnetic field at the stellar centre, $B_c$. The round, triangle and square dots have the same meaning as in Fig.~\ref{fig:mraio}. In the case of rapid transitions at the interface, stellar configurations located  to the left of the maximum mass configuration are stable and those to the right are unstable. In this case,   $\partial M / \partial \epsilon_c < 0$ is a sufficient condition for dynamic instability (as in the case of cold-catalysed one-phase stars). Configurations between the triangle dot and the square dot are also unstable.  In the case of slow transitions,  the continuous curve represents the stable stars and the dashed curve the unstable ones. Notice that to the right of the maximum mass configuration there is a stable branch that verifies $\partial M / \partial \epsilon_c < 0$. Clearly, $\partial M / \partial \epsilon_c < 0$ is no longer a sufficient criterion for dynamic instability in the case of slow conversions.} 
    \label{fig:mrho}
\end{center}
\end{figure}

The $M$--$R$ relationship for  magnetised hybrid star families  is presented in Fig.~\ref{fig:mraio}. In both panels, the maximum mass increases with the central magnetic field. Although the difference is of  $\sim 1 \%$, this monotonous increment of the maximum mass, as the MF increases up to $B_{\rm{max}} \sim 3.0\times 10^{18}$~G, is a general result, independent of the FCM parameters values. The round dot in the curves indicates the appearance of a quark matter core inside the star; which for the presented EoSs occurs close but before the maximum mass configurations. Stable configurations from this dot to smaller radii have an increasingly large quark matter core. Moving beyond the maximum mass in the direction of smaller radii, hybrid stellar configurations verify the condition $\partial M/\partial \epsilon_c < 0$ (see Fig.~\ref{fig:mrho}), and as explained in Section~\ref{sec:dynamical_stability}, they are dynamically unstable if the quark-hadron phase transition at the interface is rapid. 

Therefore, for the presented EoSs, only a few  stable stellar configurations are hybrid stars if the phase conversion is rapid.

However, if the phase conversion is slow, the branch to the left of the maximum mass, represented with a continuous curve, is also stable because the fundamental oscillation mode verifies $\omega_0^2>0$ along it. The stability of the fundamental mode is lost at the terminal configuration, which is indicated with an asterisk in Fig.~\ref{fig:mraio} (at that point, $\omega_0^2=0$). Beyond the terminal configuration, an unstable branch arises (represented with a dashed line) in which $\omega_0^2<0$. The triangle dot indicates the stellar configuration where the first radial mode becomes unstable (i.e., $\omega_0^2< \omega_1^2 <0$ beyond it) while the square dot indicates where the second radial mode becomes unstable  (i.e., $\omega_0^2< \omega_1^2 < \omega_2^2 <0$ beyond it).  Notice that, unlike the case of cold-catalysed one-phase stars,  changes of stability of the radial modes do not occur at maxima or minima in the  $M$--$R$ diagram in the case of slow conversions.

A general result we have obtained is that the length of the \textit{extended} hybrid stable branch starting at the maximum mass configuration and ending at the terminal one, shortens if the transition pressure increases. This last quantity increases both, if any of the two FCM parameters increases or if the magnetic field strength at the transition increases. Regarding the existence of this extended stable branch, we find that it appears only if the maximum mass star has already a quark core. Otherwise, and if the maximum mass star is purely hadronic, the stability of the fundamental mode is not preserved. For sufficiently large values of one of the two FCM parameters, the extended branch disappears and the last stable star is the maximum mass one, coinciding with the rapid transition criterion.

In Fig.~\ref{fig:mrho}, we present our results for the relationship among the mass, the central energy density and the central MF. As in the $M-R$ figures, stable stellar configurations in the slow transition scenario are represented by continuous curves, the rounded dots indicate the appearance of the quark phase and the triangle and square dots indicate the destabilisation of the first and second excited oscillatory radial modes. From these Figures, it is clear that in the slow transition scenario $\partial M / \partial \epsilon_c < 0$ is no longer a sufficient criterion for dynamic instability.

\subsection{Baryon mass}

So far, and every time it has not been specified, we have used the term $mass$, $M$, to refer to the NS gravitational mass, given by
\begin{equation}
\label{mgra}
M = M_G = \int_0^R 4 \pi r^{2} \, \epsilon(r) \, dr \,. 
\end{equation}
From this Subsection and onwards, we will use $M_\text{G}$ for the gravitational mass, to distinguish it from the baryon mass, $M_B$, given by
\begin{equation}
\label{mbar}
M_B = m_N \int_0^R \frac{4 \pi r^{2} \, n_B(r)}{[1 - 2 G m(r)/r]^{1/2}} \, dr \,, 
\end{equation}
where $m_N$ is the nucleon mass and $n_B(r)$ is the baryon number density.

According to \cite{Bombaci1}, when there is a dynamical evolution of stars, the concept of maximum mass arising from TOV equations for NSs is partially inadequate. For an isolated compact object, it is reasonable to consider that $M_B$ remains constant along the evolution process while $M_G$ not necessarily does. In this sense, the $M_G$--$M_B$ plane provides an adequate way to study the evolution and stability of NSs. This method has already been applied in \cite{Mariani} for the study of the fast cooling evolution of proto-HSs. In this work, we will use it instead to study the evolution of  magnetised HSs due to MF decay. It will be also employed to analyse the evolution of the magnetised HSs as a consequence of the transition between two equal-baryon mass stable configurations, due to the existence of an extended stable branch.

\begin{figure}
\begin{center}
\includegraphics[width=0.8\columnwidth]{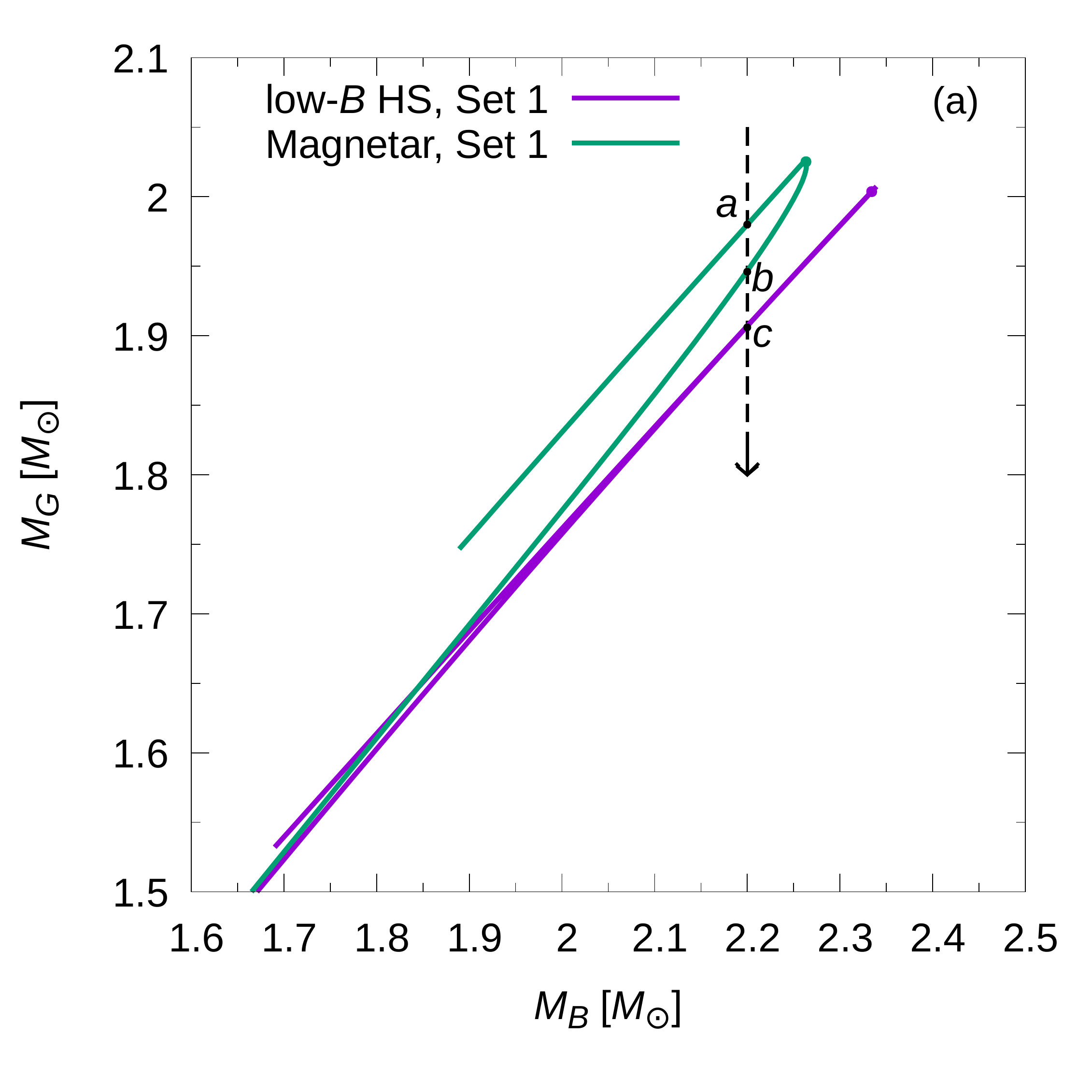}
\includegraphics[width=0.8\columnwidth]{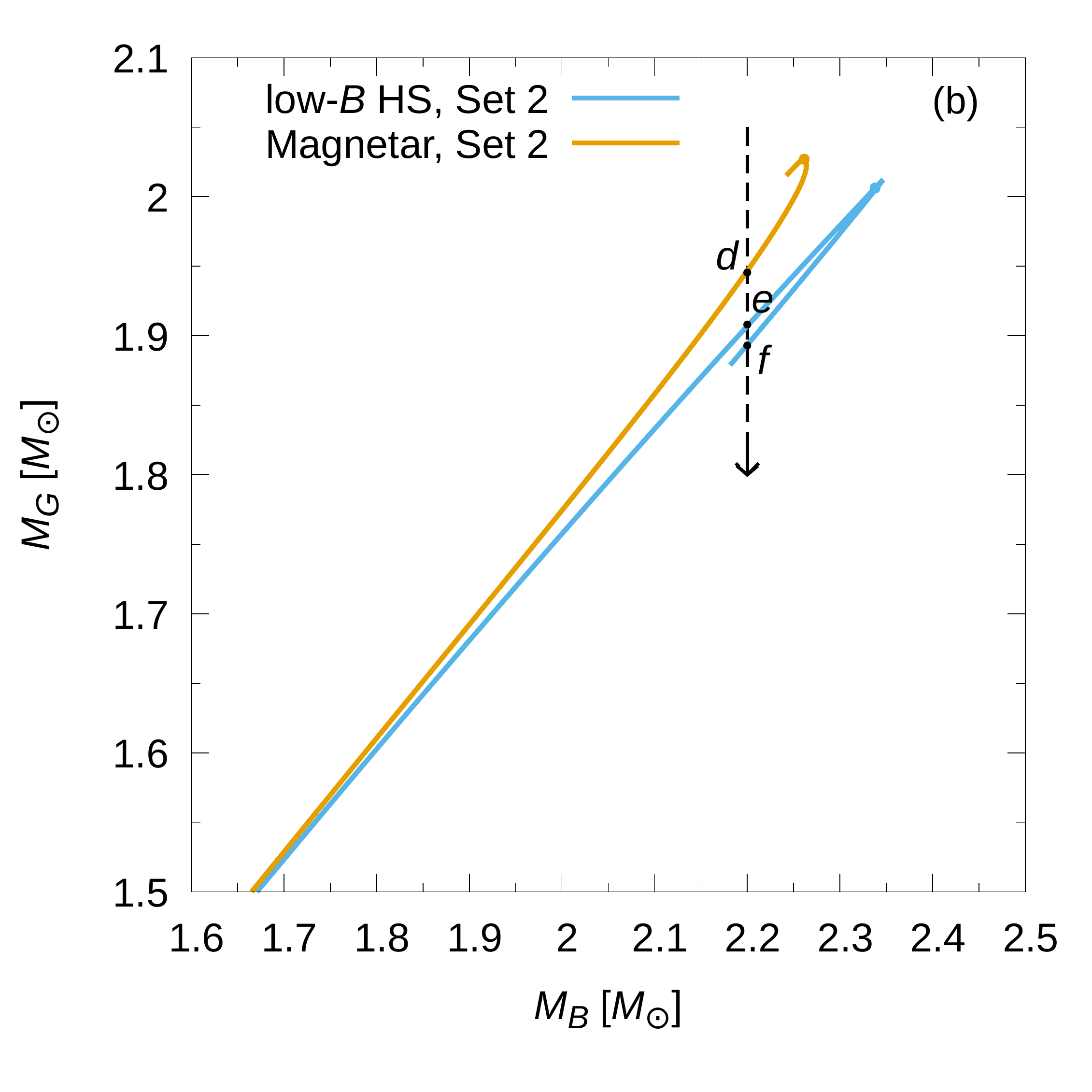}
\caption{(Colour online) $M_{G}$--$M_{B}$ plane for the two cases, \mbox{low-$B$ HSs} and magnetars, for Set 1 of the FCM parameters. In each panel, only stable stellar configurations are shown and the rounded dots indicate the appearance of the quark matter core. Considering the MF decay, from a magnetar to a \mbox{low-$B$ HS}, the star evolves along a constant baryon mass vertical line from the highest to the lowest central MF.}
\label{fig:mbmg}
\end{center}
\end{figure}

In Fig.~\ref{fig:mbmg}, we present the gravitational mass, $M_G$, as a function of the baryon mass, $M_B$, for \mbox{low-$B$ HSs} and magnetars, and, for both sets of the FCM parameters. As the MF decays, we can expect a transition from the magnetar curve to the \mbox{low-$B$ HS} curve, downwards along a vertical line, which represents the conservation of the baryon mass through the isolated evolution of the star. In Fig.~\ref{fig:mbmg}~(a) and (b) we show a vertical arrow crossing the stability curves as an example of a possible evolution, where we have labelled the involved stable configurations to analyse this process in a qualitative way. Notice that the label $c$ indicates an almost double degenerated configuration: the stable branch before the maximum mass as well as the connected extended stable branch after the maximum are nearly overlapped in the $M_G$--$M_B$ plane. Thus, the label $c$ indicates two different configurations with almost the same gravitational and baryon mass, one purely hadronic and the other one hybrid. 

Regarding the MF decay, from the first stable branch of the magnetar (configurations $a$ and $b$ in Fig.~\ref{fig:mbmg}~(a) or $d$ in Fig.~\ref{fig:mbmg}~(b)), it can be seen that  a possible transition might occur from an hybrid magnetar (configuration $a$) to a purely hadronic (configurations $c$ or $e$) or to a hybrid low-$B$ NS (configuration $f$); and it is also possible to occur a transition from a purely hadronic magnetar (configurations $b$ or $d$) to a purely hadronic low-$B$ NS (configurations $c$ or $e$) or to a hybrid low-$B$ NS (configuration $f$). This means that a phase transition could happen throughout these processes in the core of the NS, from quarks to hadrons or from hadrons to quarks,  depending on the case. It is important to remark at this point that the appearance of deconfined quark matter to form the initial hybrid magnetar (configuration $a$) could occur during the proto-NS phase \citep{Mariani}.

However, not only the slow MF decay could produce transitions between stable configurations. The existence of stable extended branches would allow transitions between stellar configurations of equal $M_B$ in the same MF-parametrisation curve. These transitions (from configuration $a$ to $b$ or from configuration $e$ to $f$ in Fig.~\ref{fig:mbmg}) could occur due to a large perturbation that destabilises the star and produces a fast catastrophic collapse event, allowing the compact object to partially expel its gravitational mass trough different forms of energy.

It is important to notice that in Fig.~\ref{fig:mbmg}, for the \mbox{low-$B$ HS} and after the maximum gravitational mass peak in the curves, the stable configurations of the extended branch have lower gravitational mass compared to the traditionally stable configuration before the maximum mass for the same baryon mass. In other words, in panel~(b), the configuration $f$ has a lower gravitational mass than the configuration $e$. In panel~(a), this phenomenon is not noticeable due to the superposition of the curves, but it is also present. This would mean that the purely hadronic low-$B$ star could collapse into a more compact HS with the occurrence of a hadron-quark phase transition if some large perturbation destabilises it.

In summary, on one hand, the study of the $M_{G}$--$M_{B}$ plane allows to consider the possibility of different channels for the MF decay and for the occurence of phase transitions. On the other hand, beyond this qualitatively analysis and the mechanisms triggering such processes, it is important to point out that if these transitions between stable configurations occur, a considerable amount of energy would be released due to the difference between the initial and final gravitational mass of the configurations involved. An estimate of such energy gives $E_{\rm{rel}} = \Delta M_G \, c^2 \sim 0.05~M_\odot \, c^2 = 5 \times 10^{52}~$erg, which is enough to explain the energetics involved in a typical GRB \citep{Nakar,Rezzolla}.

\subsection{Tidal deformability}

The NS tidal deformability, $\lambda$, is an important parameter for NS-gravitational wave astronomy since it determines the pre-merger gravitational wave signal in binary NS merger events.

Keeping only linear terms, $\lambda$ is related to the dimensionless tidal Love number, $k_2$, associated with quadrupolar $\ell =2$ perturbations,
\begin{equation}
\lambda = \frac{2}{3}k_2R^5 \, .
\end{equation}

The dimensionless tidal deformability, $\Lambda$, is defined as
\begin{equation}
\Lambda = \lambda/M^5.
\end{equation}

\begin{figure}
\begin{center}
\includegraphics[width=0.8\columnwidth]{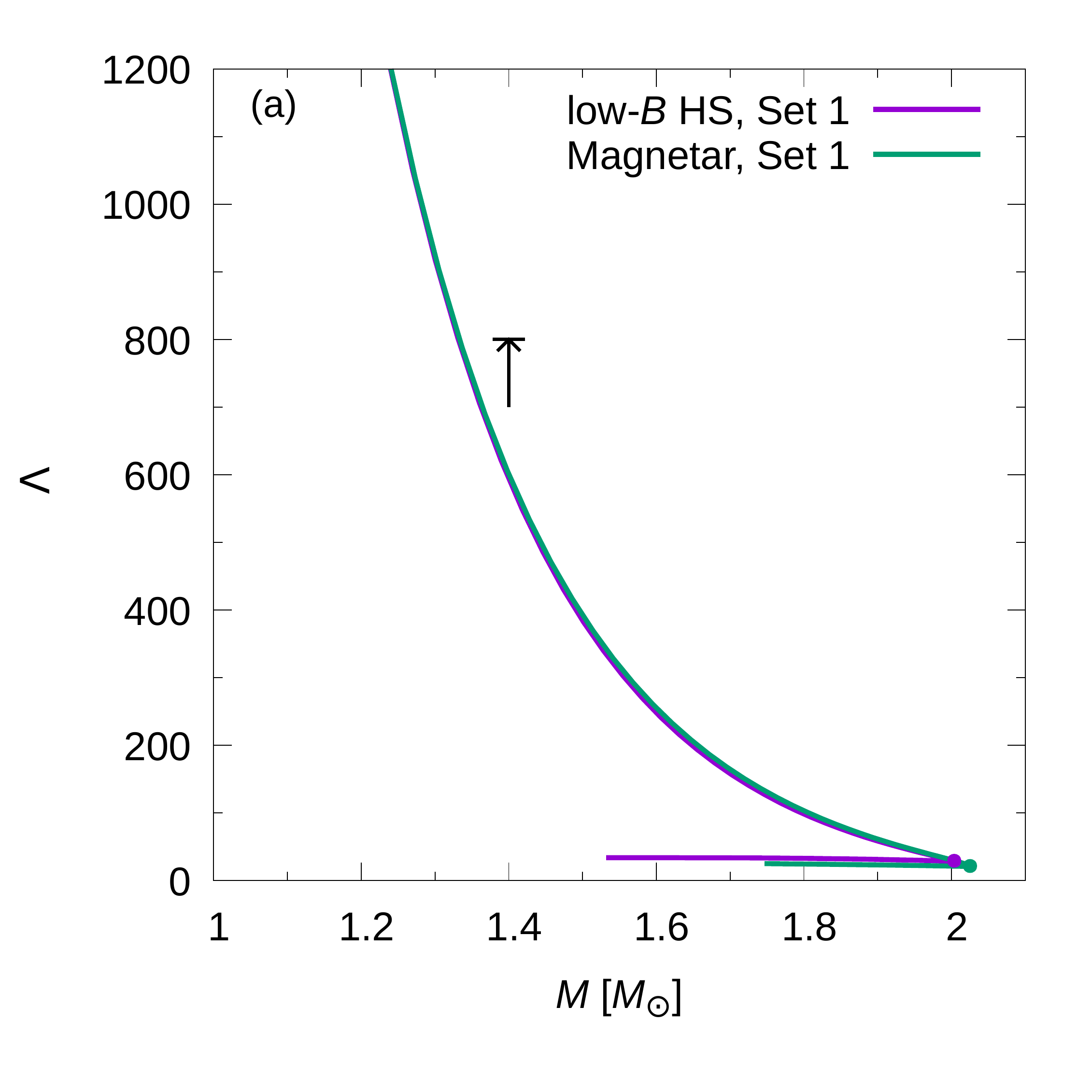}
\includegraphics[width=0.8\columnwidth]{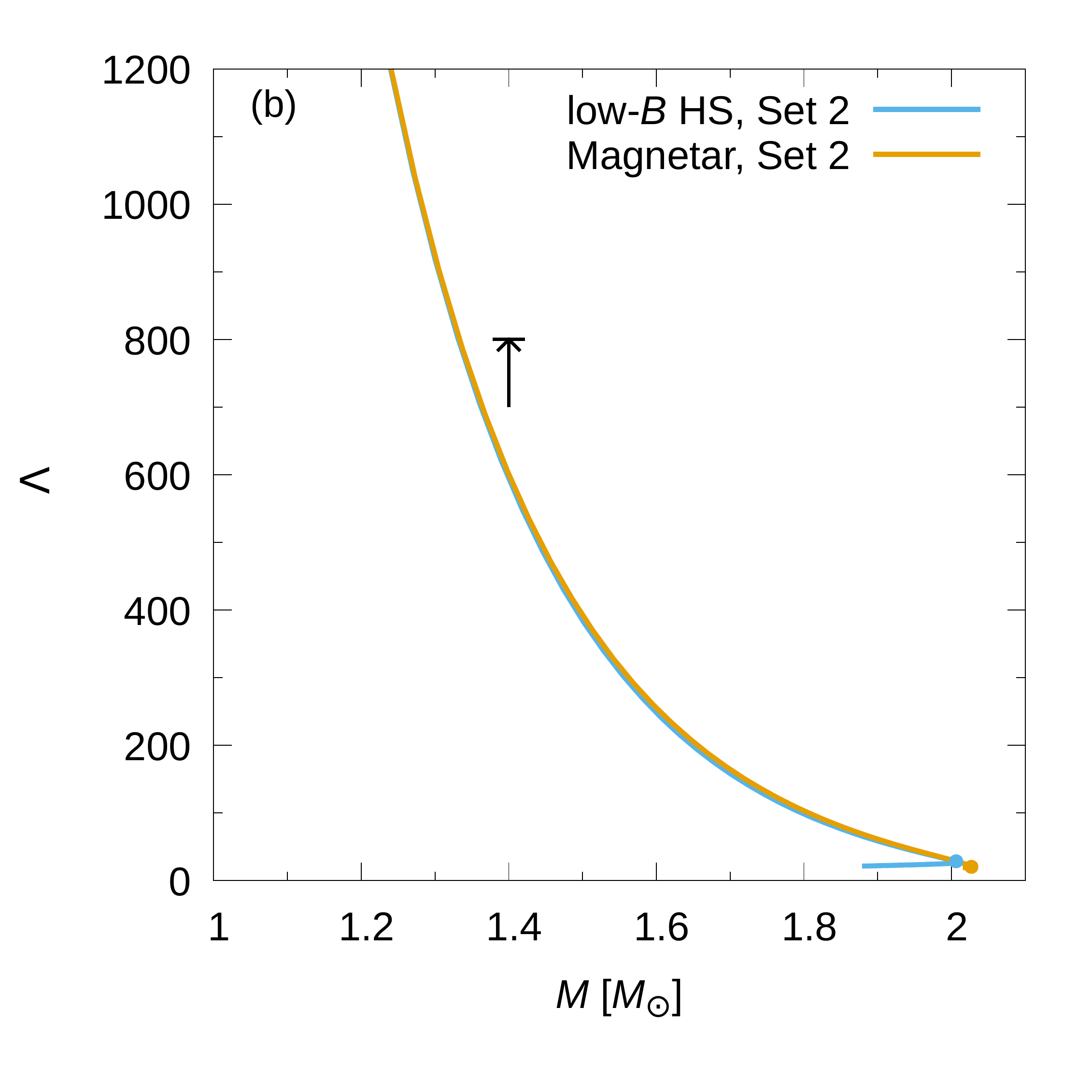}
\caption{(Colour online) Dimensionless tidal deformability, $\Lambda$--$M_G$ plane for \mbox{low-$B$ HSs} and magnetars, for Set 1 (a) and Set 2 (b) of the FCM parameters. Only stable configurations are presented and the black vertical arrow marks the constrain obtained  in \protect\cite{Abbott1} for GW170817, $\Lambda_{1.4} < 800$.}
\label{fig:tidal}
\end{center}
\end{figure}

For details regarding the calculations involved in obtaining $\Lambda$ for stellar models with sharp discontinuities in their energy density profiles see \cite{Han}.

Data analysis from GW170817 put strong constrains on the dimensionless tidal deformation parameter of a $1.4~M_\odot$ NS, $\Lambda _{1.4} \le 800$ \cite{Abbott1,Annala,Most,Raithel}. In Fig.~\ref{fig:tidal} we present the theoretical curves obtained for a \mbox{low-$B$ HS} and a magnetar, using both sets of the FCM parameters for $\Lambda$ as a function of the gravitational mass. We see that this relationship ceases to be of functional nature as there appears a degeneracy. Such degeneracies are a consequence of the appearance of extended stable branches in the case of slow conversions. In panels~(a) and (b), it can be seen that the observational constrain from GW170817, indicated by an arrow, is fulfilled.


\section{Summary, discussion and conclusions}
\label{conclus}

In this work, we have studied the effect of strong MFs in the structure and stability of HSs and the astrophysical consequences of possible transitions between different branches of stable HS configurations. We carried out this study for two paradigmatic cases: \mbox{low-$B$ HSs} and magnetars. The hybrid magnetised EoS was obtained by combining the GM1L parametrisation of the RMF model for the hadron phase and the FCM model for the quark phase. Although it has been already shown that the hybrid EoS GM1L+FCM without MF satisfies the observational mass constraint of $2 M_{\odot}$ (see, for example, \cite{Ranea,Orsaria:2019pti}), we have extended the study of hybrid stars considering the effect of the MF on the EoS describing their matter composition. Our results show that both the $2 M_{\odot}$ limit and the tidal deformability constraint are verified both when magnetized GM1L hadronic EoS or magnetised GM1L+FCM hybrid EoS are used.

In order to obtain representative general results that satisfy the $2~M_\odot$ constraint, we considered two sets of the FCM parameters. To include the MF contribution, we used an ansatz for the MF, parametrised via the baryon chemical potential, and taking into account Landau quantization. Assuming the occurence of a sharp hadron-quark phase transition, we used the Maxwell construction in which electric and baryonic charges are conserved locally. Using the NeStOR code to automate the whole process,  stellar structures were calculated. Moreover, we analysed the stability of stellar configurations considering rapid and slow phase transition scenarios.

In addition, we studied the matter magnetisation and particle population contributions for the hadron and quark phases. In both phases we obtained that the magnetisation is extremely sensitive to changes in the particle population. Moreover, the pressure due to the magnetisation (and so the anisotropies in the matter pressure) is non-negligible only for extremely high MF. As a consequence, according to the ansatz considered for the MF profile, for extremely strong magnetic fields we include only the contribution of the magnetisation pressure coming from the quark phase, since there are no more hadrons at such high densities.

On the other hand, the pure magnetic contribution to the EoS becomes highly relevant as the MF increases, leading to an increased difference between the pressure components parallel and perpendicular to the local direction of the magnetic field. As discussed in Sections~\ref{intro} and \ref{eos}, assuming that the MF in the core is  a complex and disordered combination of poloidal and toroidal fields of the same order, this anisotropy can be suitable treated via an average of the pressure components and considering that a spherical symmetric stellar configuration is a valid approximation. 

Then, using the spherically symmetric TOV equations we obtained several stellar properties such as the gravitational mass, the radius, the baryon mass, the tidal deformability and the stable branches of magnetised HSs. In Fig.~\ref{fig:mmax1}, besides the dependence of the maximum mass with the FCM parameters, we found a monotonic increase of the maximum mass of about $\sim 1\%$ when the MF increases. Furthermore, the stability of hybrid configurations depends strongly on the kind of phase transition at the sharp interface (rapid or slow).For rapid transitions, stellar configurations remain stable (i.e., have real eigenfrequencies) only if $\partial M/\partial \epsilon_c > 0$, while an extended stable branch beyond the maximum mass peak appears if the transition is slow. At the extended stable branch, $\partial M/\partial \epsilon_c < 0$ is verified; however, in spite of this, all radial eigenfrequencies are real indicating that radial perturbations do not grow with time. For the set of EoSs adopted in this work, a pure quark matter core arises for stellar configurations that are close to that with the maximum mass star (but before it). Since, in the scenario of rapid transitions, the terminal mass coincides with the maximum mass, there is a very narrow range of central densities that lead to stable hybrid stars in this case. On the other hand, in the scenario of slow transitions, the terminal configuration occurs beyond the maximum mass one, meaning that a wider family of HSs is possible. In fact, the central density in the slow case can reach values considerably larger than in the rapid case. Finally, it is worth noticing that within the slow transition hypothesis, the central densities for stabilisation/destabilisation of the normal modes are shifted with respect to the ones predicted in the rapid transition scenario, $\partial M/\partial \epsilon_c = 0$. A more detailed analysis of this result is left for a future work.

Not only the MF strength but also the FCM parameters affect the length of the extended stable branch: the larger the parameter values are, the shorter is the length of such branch. Furthermore, we have explored other parameter combinations of the FCM. For slow conversions, we have found that, if the parameters values are such that quark matter appears only for configurations beyond the maximum mass star, then the fundamental mode has zero frequency at the maximum mass and the extended stable branch does not exist. In these cases, the terminal mass coincides with the last stable star for the rapid transition stability criterion. For the particular case of the terminal mass configurations, a baryon number density up to $20$ times the nuclear saturation density, $n_0 \approx 0.16$~fm$^{-3}$, is reached in their centre. This result should motivate the research of less explored and more extreme regions of the QCD phase diagram.

On the other hand, an analysis of the $M_G$--$M_B$ plane allowed us to study possible evolutionary paths of the HS as the MF decays. In this sense, we presented a qualitative analysis of different channels through which magnetars can evolve, including the possibility of phase transitions. It has been suggested that a phase transition from hadron to quark matter could occur in the core of a recently born NS \citep{Mariani,Prakash,Benvenuto}. However, the results in this work suggest that this phase transition could also occur during the slow MF decay expected to occur when a magnetar evolves to become a low-B object. Within this scenario, quark matter would appear not at the proto-NS stage but probably thousands of years after the NS brith. Additionally, a transition between two stable configurations could occur if a sufficiently strong external perturbation destabilizes a star located along the upper branch of a given equilibrium sequence in Fig. \ref{fig:mbmg}, triggering a fast collapse to the lower branch of the same sequence.

We have calculated the energy released in  transitions between stable stellar configurations with the same baryon mass and found that it is $E_{\rm rel} \sim 5 \times 10^{52}$~erg, which would be enough to power a GRB event.  In fact, the idea that the hadron-quark conversion could work as the central engine of at least some GRBs, has been  proposed several years ago \citep{Olinto:1986otc,Cheng:1996con,Xu1999:gas}. Many subsequent works have shown that this scenario is energetically feasible,  involving an energy release of about $\sim 10^{52}-10^{53}$~erg \citep{Bombaci:2000con,Ouyed:2002qsa,Berezhiani:2003grb,Marquez:2017pti}. The main new ingredient of the present study is that one of the configurations involved in the transition (either the initial or the final one) pertains to the new stable branch.  In spite of this, the energy scale is roughly the same in all the above scenarios. Additionally, some works have focused on possible mechanisms  that would convert the available energy into pure gamma radiation.  For example,  \citet{Haensel:1991um, Cheng:1996con,Lugones:2002vj} argue that a small fraction of the neutrinos and antineutrinos, produced in the stellar core during the phase conversion, may annihilate into electrons and positrons above the compact star surface giving rise to an expanding fireball (mixture of photons, electrons, and positrons) that could explain the observed gamma ray emission. Some clues for beaming the radiation in magnetized compact stars have also been provided \citep{Lugones:2002vj}. There are also studies of the implications of quark deconfinement on the phenomenology of long gamma-ray bursts focusing, in particular, on the possibility to describe multiple prompt emission phases in the context of the proto-magnetar model \citep{Pili2016a}.

Last but not least, the recent new constrains arising from the gravitational wave detections were considered. In the $M$--$R$ plane, we verified that the R$_{1.4} \lesssim 13.76 \, \mathrm{km}$ condition \citep{Raithel,Annala,Malik,Most,Fattoyev}  is satisfied together with the restriction of  $\tilde{\Lambda} < 800$ \citep{Abbott1} in the $\Lambda$--$M$ plane.

Regarding the degeneracy obtained in our theoretical curve, it is important to point out that Advanced LIGO is expected to detect up to 50 BNS mergers during O3 run (10 is a less optimistic number) \citep{LIGO}. This amount of data would be extremely useful to get more insight about the composition of NSs. If one of the components of the  binary system of such events had a sufficiently large mass (in the range $\sim 1.6 - 2.0 \,M_\odot$ according to Fig. \ref{fig:tidal}), the existence of an extended branch of stable NSs could be tested. However, it should be noted that optimism about detecting such objects should not be overstated since observed NSs in binary systems have a narrow mass distribution with a peak at $1.33~M_\odot$ and a dispersion of $0.05~M_\odot$ \citep{Ozel_2012} (similar results can be found in \cite{Piro_2017}).

The results presented in this work were obtained in a context in which  multi-messenger astronomy promises new observations in the short term, impossing more constraints and challenging both, the current micro and macro physical models. Regarding  dense matter models, the uncertainties about whether a phase transition between hadron and quark matter could occur at high densities and low temperatures, as well as its nature (sharp or soft, rapid or slow), should be addressed with new perspectives. From the astrophysical point of view, the gravitational wave asteroseismology of NSs, through the calculation of non-radial oscillation modes, is a new promising tool that, with future GW detector generations, may open a new window to study these compact objects. We plan to address the above issues in future works.

\section*{Acknowledgements}

The authors thank the anonymous referee for the constructive comments and criticisms that have contributed to improve the manuscript substantially. MM, MGO and IFR-S acknowledge support from Universidad Nacional de La Plata and CONICET under grants number G140, G157 and PIP-0714. GL is thankful to the Brazilian agency Conselho Nacional de Desenvolvimento Cient\'{\i}fico e Tecnol\'{o}gico (CNPq) for financial support.



\bibliographystyle{mnras}
\bibliography{mnras_mariani} 



\appendix

\section{Units}
\label{appe}

\subsection{Electromagnetic units} 
\label{appe1}

In this work we use the Heaviside-Lorentz (HL) system of units. This is a rationalized system within cgs that takes $\epsilon_0 = \mu_0 =1$, where $\epsilon_0$ is the vacuum electric permittivity and  $\mu_0$ is the vacuum magnetic permeability. Within this system of units, it is important to remark that each particular electromagnetic quantity has a definite expression to keep the consistence of the system. In order to clarify this, we follow \cite{Heras} where a method for tranformations between different unit systems (SI, Gaussian and natural HL) is developed.

In the HL system of units, the electromagnetic lagrangian is given by
\begin{equation}
	{\cal L}_{_{EM}}= -\frac{1}{4} F_{\mu\nu}F^{\mu\nu} \, ,
\end{equation}
where $F^{\mu\nu}$ is the electromagnetic field tensor; 
and the electromagnetic energy density is
\begin{equation}
	u_{_{EM}}= \frac{1}{2} (E^2 + B^2) \, .
\end{equation}
Finally, considering also that we use a natural unit system where $c=\hbar=1$, it is possible to obtain the unit of the elementary electric charge via the fine-structure constant definition,
\begin{equation}
	e = \sqrt{4\pi \alpha} \, , \ \text{with} \ \alpha \sim 1/137 \, .
\end{equation}

\subsection{Magnetic Field units}
\label{appe2}

In this work we use two different units of reference for the magnetic field: Gauss and MeV$^2$. The Gauss unit is mostly related to  astrophysical references. On the other hand, the MeV unit is used in theoretical models of nuclear physics.

Although it is not entirely clear in the literature how to transform between Gauss and MeV$^2$, it is possible to obtain a relation following the definition of the magnetic field critical value for the electron \citep{Broderick},
\begin{equation}
	B_c^e = \frac{\hbar c}{e} \lambdabar_e^{-2} \, ,
\end{equation}
where $\lambdabar_e = \hbar/m_e c$ is the Compton wavelength of the electron.

In the cgs system, $B_c^e = 4.414 \times 10^{13} \, \mathrm{G}$. In the natural Heaviside-Lorentz system we use, considering $c = \hbar = 1$, $m_e = 0.511~\text{MeV}$ and $e = \sqrt{4\pi \alpha}$ as we established in Appendix~\ref{appe1}, the critical field turns out to be $B_c^e = 0.862~\text{MeV}^2$.

Comparing these two results for the magnetic critical field we obtain that
\begin{equation}
	1~\text{G} = 1.952 \times 10^{-14}~\text{MeV}^2 \, .
\end{equation}


\bsp	
\label{lastpage}
\end{document}